\documentclass[preprint,12pt,authoryear]{elsarticle}

\usepackage{amssymb}
\usepackage{amsmath}
\usepackage{mathrsfs}
\usepackage{color}
\usepackage{hyperref}
\usepackage{booktabs}
\usepackage{soul}
\usepackage{blindtext, pdflscape}
\usepackage{changepage}

\newcommand\rev[1]{\textcolor{black}{#1}}

\usepackage{lineno}

\journal{Icarus}

\begin{document}

\begin{frontmatter}



\title{Global climate modeling of Saturn's atmosphere.\\ Part III: Global statistical picture of zonostrophic turbulence in high-resolution 3D-turbulent simulations}




\author[1]{Simon Cabanes\corref{cor1}}
\author[1,2]{Aymeric Spiga}
\author[1,3]{Roland M. B. Young}

\cortext[cor1]{Corresponding author. Email address: cabanes.simon@gmail.com}

\address[1]{Laboratoire de M\'{e}t\'{e}orologie Dynamique (LMD/IPSL), Sorbonne Universit\'{e}, Centre National de la Recherche Scientifique, \'{E}cole Polytechnique, \'{E}cole Normale Sup\'{e}rieure, Paris, France.}
\address[2]{\rev{Institut Universitaire de France (IUF), Paris, France}}
\address[3]{\rev{Department of Physics \& National Space Science and Technology Center, UAE University, Al Ain, United Arab Emirates}}


\begin{abstract}
We conduct an in-depth analysis of statistical flow properties calculated from the reference high-resolution Saturn simulation obtained by global climate modelling in Part II. In the steady state of this reference simulation, strongly energetic, zonally dominated, large-scale structures emerge, which scale with the Rhines scale. Spectral analysis reveals a strong anisotropy in the kinetic energy spectra, consistent with the zonostrophic turbulent flow regime. By computing spectral energy and enstrophy fluxes we confirm the existence of a double cascade scenario related to 2D-turbulent theory. To diagnose the relevant 3D dynamical mechanisms in Saturn's turbulent atmosphere, we run a set of four simulations using an idealized version of our Global Climate Model devoid of radiative transfer, with a well-defined Taylor-Green forcing and over several rotation rates (4, 1, 0.5, and 0.25 times Saturn's rotation rate). This allows us to identify dynamics in three distinctive inertial ranges: (1) a ``residual-dominated'' range, in which non-axisymmetric structures dominate with a $-5/3$ spectral slope; (2) a ``zonostrophic inertial'' range, dominated by axisymmetric jets and characterized by the pile-up of strong zonal modes with a steeper, nearly $-3$, spectral slope; and (3) a ``large-scale'' range, beyond Rhines' typical length scale, in which the reference Saturn simulation and our idealized simulations differ. In the latter range, the dynamics is dominated by long-lived zonal modes 2 and 3 when a Saturn-like seasonal forcing is considered (reference simulation), and a steep energetic decrease with the idealized Taylor-Green forcing. Finally, instantaneous spectral fluxes show the coexistence of upscale and downscale enstrophy/energy transfers at large scales, specific to the regime of zonostrophic turbulence in a 3D atmosphere.

\end{abstract}



\end{frontmatter}


\section{Introduction}

The self-organization of large-scale coherent flows is ubiquitous in nature and is an essential, cutting edge research topic in fundamental fluid dynamics, geophysics and planetary sciences \citep{galperin19}. The strong zonal (i.e. east-west) winds observed on the gas giants, i.e. Jupiter, Saturn, Uranus and Neptune \citep{Inge:90} are unmistakable markers of powerful, self-organized, atmospheric dynamics, insofar as  they have been difficult to access by direct measurements. 

By tracking cloud motions in the upper atmosphere, ground-based observations and the successive Voyager, Cassini, New Horizons, and Juno missions have provided detailed information on the wind patterns of Jupiter and Saturn, including temporal variability and vertical profiling \citep[e.g.,][]{porco03,sanchez04,vasavada05,garcia10}. Each of these planets exhibits a system of axisymmetric (i.e. axial symmetry relative to the rotational spin axis) jet-streams, confined in latitude and almost invariant in azimuth. These jets are latitudinally-varying winds that alternate between prograde (eastward) and retrograde (westward) flow with respect to the apparent mean planetary rotation. Latitudinal jet profiles exhibit qualitative similarities on both Jupiter and Saturn: there exists a broad, fast prograde equatorial jet and numerous narrower, weaker prograde jets at higher latitudes. Despite these similarities, the zonal jets are considerably broader, faster, and fewer in number on Saturn than on Jupiter \citep{showman17}. 

Apart from the dominant axisymmetric flow, hundreds of non-axisymmetric eddies develop (convective storms, waves, vortices, and turbulence). Jupiter's Great Red Spot, an anticyclonic vortex \citep{Flet:10grs,Simo:14}, and Saturn's polar hexagon, a wavenumber-6 disturbance superimposed on the northern circumpolar jet \citep{Bain:09,Sanc:14hexagon,Rost:17saturn} constitute the largest and the most long-lived examples. In recent decades, much effort has been made by theoreticians, experimentalists, and numericists to understand the relevant dynamical mechanisms responsible for the axisymmetric jets, the non-axisymmetric eddies, and the interactions between them \citep{Saly:06,Delg:12,galperin14}.

How do axisymmetric jets emerge in gas giants? It is well accepted that in three-dimensional (3D) systems, turbulent kinetic energy is typically transferred from larger to smaller scales through a direct energy cascade (the Kolmogorov cascade). However, this process can be reversed under certain conditions favoring coherent flow. In other words, non-axisymmetric turbulent flows \citep[e.g. caused by flow baroclinicity,][]{salmon80} can feed the large-scale jets. This is the case in both quasi-two-dimensional (2D) domains, where the flow is geometrically confined in depth to a shallow layer, or under rapid rotation in 3D domains, where the flow becomes invariant along the rotational spin axis and is dynamically confined to quasi-2D geostrophy. \rev{Two types of models have been proposed for jet formation in planetary atmospheres: shallow models where jet forcing is confined to the atmospheric weather layer \citep{cho96,lian08,schneider09,liu10,Youn:19partone,spiga2020} and deep models where jet forcing extends into the convective molecular envelope of the planet \citep{christensen02,heimpel05,gastine14,cabanes17}}. In both cases, flow bi-dimensionalization transfers kinetic energy from small to large scales via the so-called inverse energy cascade \citep{Kraichnan67}. \rev{It is only recently that this theoretical framework has been generalized to 3D rotating turbulence by \citet{sukoriansky16}. Using analytical tools, they showed the transition from a 3D direct energy cascade to rotation-dominated (quasi-2D) turbulence leading to an inverse energy cascade.} 
This duality between \rev{2D and 3D turbulence as well as} geometrical and dynamical flow confinement, \rev{together with the associated transition from direct to inverse energetic cascade,} is at the core of fundamental geophysical fluid dynamics \citep{smith96, pouquet2013, deusebio14}.

Spherical geometry also carries an additional background potential vorticity gradient $\beta$, which preferentially channels kinetic energy in the zonal direction, promoting large-scale zonal motions \citep{vallis93,cho96}. The planetary vorticity $\beta=(2\Omega/a)\cos\varphi$ directly results from the variation of the Coriolis parameter with latitude $\varphi$, the rotation rate $\Omega$, and planetary radius $a$.
This coupled action of the turbulent inverse cascade and of the $\beta$-effect in planetary atmospheric envelopes links large-scale features to smaller scale dynamics through the continuous process of scale-to-scale interactions. These specific planetary conditions define the regime of zonostrophic turbulence, coined by \cite{sukoriansky02}. 

In the framework of 2D-turbulence, \cite{sukoriansky02} generalized the double cascade scenario of Kolmogorov-Batchelor-Kraichnan (KBK) (i.e. the inverse cascade of energy and the direct cascade of enstrophy) to turbulence on the surface of a rotating sphere.  They showed that 2D-turbulent zonostrophic regimes display universal kinetic energy spectra with a steep $-5$ slope in their axisymmetric component, while in the non-axisymmetric component the classical KBK scaling prevails with a $-5/3$ slope. This spectral anisotropy, as well as the presence of an inverse energy cascade, has been detected in Jupiter's tropospheric winds using Cassini near-IR imaging \citep{choi11,galperin14,young17}. 

Similarly, this zonostrophic signature has been identified in 2D-turbulent numerical simulations both with a $\beta$-plane approximation \citep{chekhlov96} and on the surface of a rotating sphere \citep{huang01, galperin06}.  These models reproduce the basic dynamical features of generic gas giant atmospheres without reproducing specific planets, yet they lack the physical realism of real planetary systems as they explicitly assume 2D-turbulent flows. The emergence of Global Climate Models (GCMs) 
for gas giants \citep{Dowl:06,schneider09, liu10, lian10,Youn:19partone,spiga2020}, where a thin planetary shell is simulated by solving for 3D primitive equations of motion with increasingly realistic physical parametrizations \citep[e.g., radiative transfer,][]{guerlet14,spiga2020}, now allows for in-depth spectral analysis of simulated giant planets in order to diagnose the relevant dynamical mechanisms in 3D-turbulent zonostrophic flow.

To step in this direction, the present paper (Part III of this series) uses high-resolution multi-annual 3D numerical simulations obtained by the Saturn DYNAMICO GCM described by \citet{spiga2020} (Part II of this series). This GCM is designed to explore Saturn's tropospheric and stratospheric dynamics with a new icosahedral dynamical core \citep[DYNAMICO,][]{dubos15} and realistic radiative transfer \citep[][Part I of this series]{guerlet14}. Following the approach that \cite{augier13} applied to terrestrial Global Climate Models, we compute kinetic energy spectra as well as spectral energy and enstrophy fluxes from instantaneous wind maps. This is done in two stages. First, we compute the spectral analysis of the Saturn reference simulation (SRS in the text below) 
described in Part II by \citet{spiga2020}, where the forcing gives rise to banded jets resulting from the growth of baroclinic eddies at small scales. Second, we run a set of idealized simulations, using only the DYNAMICO dynamical core, with a well-defined Taylor-Green forcing but with radiative transfer being switched off, which suppresses flow baroclinicity caused by equator-to-pole gradients. In these idealized simulations we vary the planetary rotation rate \citep[as done by, e.g.,][]{Chem:15} to bracket the properties of Saturn's zonostrophic regime, and to diagnose the relevant dynamical processes in a 3D-turbulent atmospheric layer without the effects of intermittency and spatial anisotropy induced by baroclinic forcing. 

In all these simulations we identify the double cascade predicted by KBK for 2D-turbulence, as well as zonostrophic spectral anisotropy. In the SRS, semi-annual and annual energy cycles are identified. Our analysis reveals that the dynamics exhibit a transition across the Rhines scale $L_R$ \citep{rhines75}, showing a large-scale dynamical regime ($L>L_R$) dominated by long-lived energetic modes, and a small-scale dynamical regime ($L<L_R$) that is highly sensitive to the seasonal cycle. In turn, the idealized simulations emphasize the peculiar dynamics that develop within the zonostrophic inertial range, where axisymmetric kinetic energy substantially dominates over non-axisymmetric kinetic energy. As the rotation rate increases, the dynamical confinement (i.e. quasi-2D geostrophy) as well as the inverse energy cascade become more efficient, leading towards a steeper energy spectra to which the classical picture of 2D-turbulence is not applicable.

In Section~\ref{Numerical setup} we remind the reader about the key characteristics of the SRS (Saturn reference simulation) and we present the numerical setup used to run our idealized simulations. The theoretical framework of zonostrophic turbulence is introduced in Section~\ref{Scaling in 3D zonostrophic turbulence} as well as the typical scaling characteristics of 3D rotating turbulence. The statistical tools are detailed in Section~\ref{Statistical analysis}. Statistical analysis is first applied to the SRS in Section~\ref{reference simulation}, then to our idealized setup in Section~\ref{Idealized simulations}. Conclusions and perspectives are in Section~\ref{conclu}.

\section{Numerical setup}\label{Numerical setup}

The first part of our analysis uses the exact same SRS as is described and analyzed in detail in \citet{spiga2020}. Here we remind the reader about the main characteristics of this simulation.  In the present manuscript, latitude is denoted as $\varphi$, longitude as $\lambda$, radial/vertical as $z$ and $g=10.4$\,m\,s$^{-2}$ is the gravitational acceleration on Saturn. We use Saturn's radius $a=58232000$\,m and rotation rate $\Omega = 1.65121 \times 10^{-4}$\,s$^{-1}$.

A Global Climate Model (GCM) consists of coupling a generic hydrodynamical solver (usually named the dynamical core) with physical parametrizations specific to a planet, aimed at emulating the processes unresolved by the dynamical core (radiative transfer, small-scale mixing, cloud formation, etc. \ldots). Developed at Laboratoire de M\'et\'eorologie Dynamique, DYNAMICO is a new dynamical core that solves the primitive equations of motion with a hybrid mass-based vertical coordinate. Equations are solved in the limit of hydrostatic dynamics with the shallow-atmosphere approximation (i.e. reduced Coriolis parameter). Spatial discretizations are formulated following an energy-conserving three-dimensional Hamiltonian approach \citep{dubos15}. The DYNAMICO dynamical core uses a quasi-uniform icosahedral mapping of the planetary sphere to ensure excellent conservation and scalability properties in a massively parallel configuration \citep{dubos15} with the projection from the unstructured icosahedral grid to a cylindrical latitude/longitude grid performed at runtime \citep[see][for details and references]{spiga2020}. This approach enables us to carry out computationally-intensive simulations combining high horizontal resolutions (here $1/2^{\circ}$ in latitude and longitude) with a long integration time of more than ten Saturn years. Time integration is done by an explicit Runge-Kutta scheme. 

Subgrid-scale dissipation in the horizontal is solved by adding a hyperdiffusion term to the vorticity, divergence and temperature equations: 
\begin{linenomath*}\begin{equation}
\left[ \frac{d \mathcal{F} }{d t} \right]_{dissip} = \frac{(-1)^{q+1}}{\tau} \nabla^{2q} \mathcal{F}
\end{equation}\end{linenomath*}
\noindent where $\mathcal{F}$ can be divergence, vorticity, or potential temperature, $q$ is the order of dissipation, and $\tau$ the damping timescale at the smallest horizontal scale resolved by the model. The selection of appropriate values for $q$ and $\tau$ and their impact on the global dynamics remains empirical, as discussed in the appendices of \citet{spiga2020}. For our idealized simulations described in section~\ref{Idealized simulations}, we choose the same values as what the ones used by \citet{spiga2020} for the SRS, i.e. $q=2$ and $\tau=10000$. Setting these hyperdiffusion parameters is a trade-off between the stability of the model integration and the degree to which small-scale perturbations are resolved. 

In the vertical dimension (i.e. along radial coordinates, which correspond to the vertical in our shallow-atmosphere simulations), both the SRS in \citet{spiga2020}, and the idealized GCM simulations presented in section~\ref{Idealized simulations}, solve the equations at 32 levels, from $3$ bar to $1$ mbar, from the upper troposphere to the middle stratosphere. 
A stress-free boundary condition is imposed at the uppermost layer, instead of an absorbing sponge layer that may result in spurious effects due to non-conservation of angular momentum \citep{shaw07}. At the lowest layer (i.e. at the 3-bar level), the bounding condition is stress-free and a bottom drag $u_b=-\alpha u$ is imposed with a frictional rate $\alpha = 1/\tau_R$ similar to \citet{schneider09} and~\citet{liu10}. This large-scale frictional dissipation emulates \rev{to zeroth order, at the bottom of our model,} the magneto-hydrodynamic drag (MHD) acting on jets \rev{extending} at the bottom of the atmospheric layer \citep{liu08} several thousand kilometers below the visible cloud layer \citep{Gala:19}. Improving on this admittedly simplistic bottom boundary condition is an entire research goal on its own and is left for future investigations. In our model, we set a frictional region with $\tau_R = $ 90 Earth days, with a non-frictional region near the equator  $-33^{\circ} < \varphi < 33^{\circ}$ with $\tau_R = \infty$, as in \cite{schneider09, liu10}. In the vertical, subgrid-scale dissipation is handled by a combination of a Mellor-Yamada diffusion scheme for small-scale turbulence, and a dry convective adjustment scheme for organized turbulence (e.g. large-scale convective plumes).

The physical parametrizations used in our GCM have been specifically developed to simulate the physical properties of Saturn's troposphere and stratosphere, in particular a seasonal radiative transfer model optimized for Saturn, complete with internal heat flux and Saturn's ring shadowing, which has been validated using available measurements, as detailed by \citet{guerlet14} and \citet{spiga2020}. This is the key setting on which the SRS and the idealized simulations presented in section~\ref{Idealized simulations} differ. We note that the current version of the model does not parametrize moist convection, which occurs in giant planets' atmospheres \citep{Gier:00} and carries substantial energy into the jet system \citep{Saya:07}. The model also inhibits large vertical gradients of potential temperature (i.e. along a planetary radius) by linearizing deep convection around a mean vertical temperature profile (the convective adjustment scheme). These physical limits, together with the shallow-atmosphere approximation, are possibly responsible for some discrepancies between direct planetary observations and DYNAMICO simulated wind fields in a Saturn-like configuration like the SRS \citep[see also section 4.4 in][]{spiga2020}. Future implementations of our Saturn DYNAMICO model will address these issues. Aware of those limitations, the present work mainly focuses on DYNAMICO's dynamical core and its capability to simulate the zonostrophic inertial range relevant for the formation of jets. 

\section{Scaling and kinetic energy distribution in 3D zonostrophic turbulence}\label{Scaling in 3D zonostrophic turbulence}

Global Climate Models are powerful tools for modelling the complexity of planetary atmospheres in a realistic fashion. The primitive equations of motion are solved in a 3D-spherical domain, for a divergent, stratified flow in which energy is provided by the seasonal cycle of incoming sunlight (all planets) and internal heat flux (giant planets). Quasi-steady jets arise in GCM simulations after several simulated Saturn years as a result of the non-linear dynamics resolved by the model. In order to diagnose these mechanisms it is relevant to analyze the kinetic energy distribution in Fourier space. Spectral analysis has the advantage that each dynamical step leading to the formation of equilibrated jets can be interpreted in terms of typical length scales or, equivalently, wavenumbers.

\begin{figure}
  \centering
    \includegraphics[width=1\textwidth,clip,viewport=25 0 660 370]{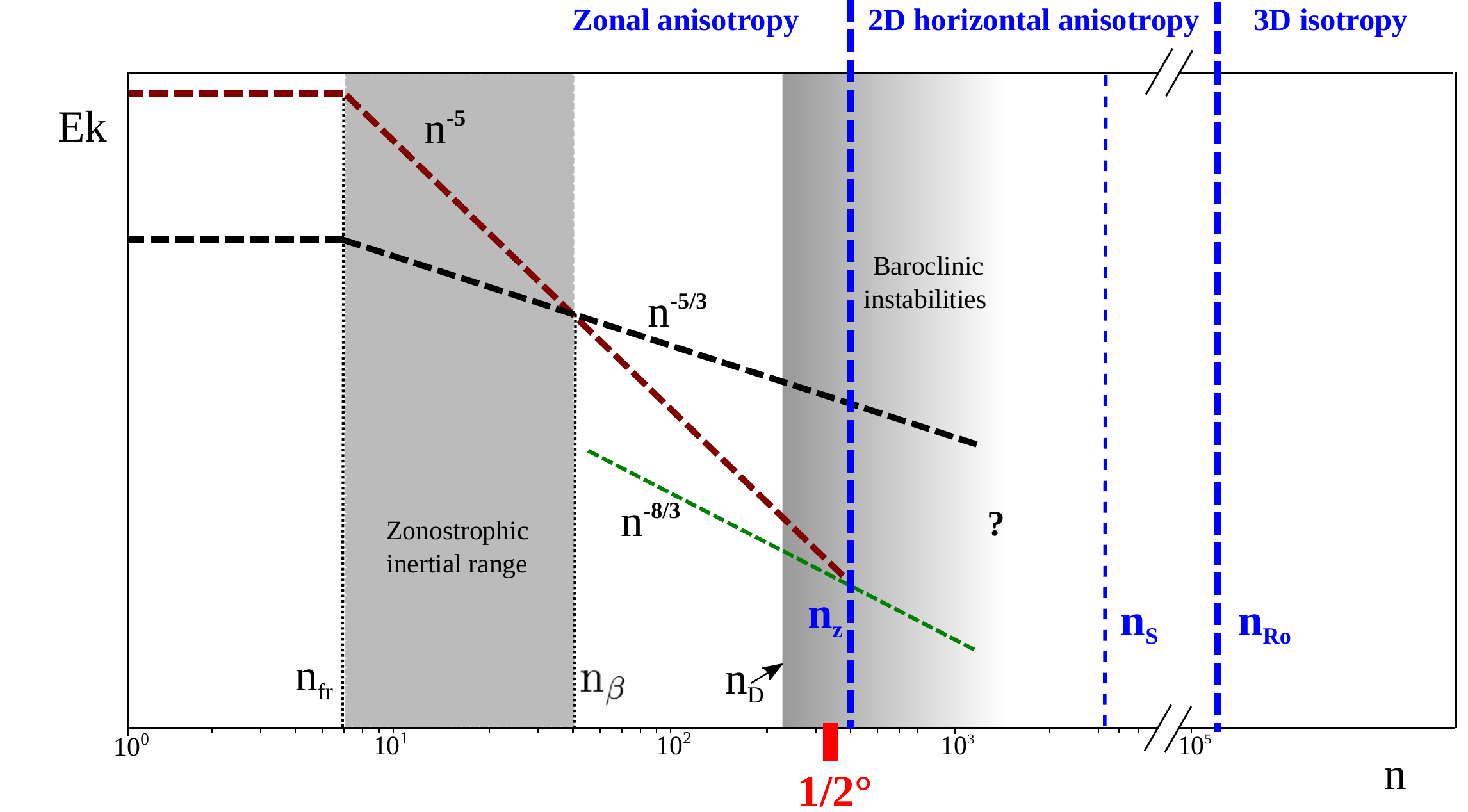}
      \caption{Theoretical energy distribution as a function of the total index $n$ \rev{presented in a similar fashion to \cite{galperin10} and \citet[][Chapter 13]{galperin19b}}. The scaling corresponds to our Saturn reference simulation. Notations are described \emph{in extenso} in the text of sections~\ref{sec:scales} and~\ref{Spectral energy budget}.\label{theory}}
\end{figure}

When spherical harmonic functions are invoked to compute energy spectra, the basis functions $Y_n^m (\varphi, \lambda)$ are eigenfunctions of the horizontal Laplacian operator $ \nabla_h^2 Y_n^m = -n(n+1) Y_n^m/a^2$, and depend on non-dimensional total and zonal indices $n$ and $m$ respectively \citep{boer83,boershepherd83}. 
The planetary spin axis defines the zonal index $m=0$ to be the axisymmetric mode that characterizes the zonal jets. Indices $m \neq 0$ define the residual non-axisymmetric part of the flow (eddies). At any typical length scale in latitude $L$, one can attribute a typical total index $n$ given by the relation $n = 2 \pi a/L$, since $\varphi \in [0, \pi]$. 
We represent in Fig.~\ref{theory} the theoretical distribution of kinetic energy as a function of total index $n$ \rev{in the same fashion as presented in \citet{galperin10}}. This diagnostic aims to give a global picture of the zonostrophic regime with a scaling suitable for our Saturn reference simulation.  This scaling is now described.

\subsection{Typical scales and the 2D horizontal anisotropy \label{sec:scales}}
The absorption of sunlight in a stably stratified, radiative, motionless atmosphere leads to the development of an unstable meridional temperature gradient, which drives small-scale baroclinic eddies. These eddies result from the conversion of potential energy, stored in density fronts, into kinetic energy at typical horizontal (i.e. along the spherical surface) length scales close to Rossby radii of deformation, $L_D = N_{BV}H/d\pi f$, where $d$ are positive integers defining baroclinic modes, $N_{BV}=\sqrt{\frac{-g}{\rho} \nabla \rho} $ is the Brunt-V\"ais\"al\"a frequency, and $\nabla \rho$ is the density change over depth $H$. 
Thus, in SRS, kinetic energy is presumably created at the typical baroclinic index $n_D \geq 2\pi a/L_D$, corresponding to the shadowed area in Fig.~\ref{theory}. In this figure, we also locate the smallest resolved scale corresponding to $1/2^{\circ}$ in latitude; this indicates that only the first Rossby deformation radius is resolved in our simulations.\\

It is commonly argued that rapid planetary rotation ``barotropizes'' baroclinic eddies, leading to an inverse kinetic energy cascade towards larger scales \citep{charney71,rhines75,salmon80}. \rev{Such a picture of energy source and energy transfer has been observed in terrestrial atmospheric and oceanic reanalysis data \citep{chemke16a,chemke16b} as well as in idealised GCM simulations \citep{chemke15}.} 
In the specific configuration of shallow-rotating atmospheric layer, dynamical confinement (i.e. barotropization) and geometrical confinement (i.e. weak depth extension) act together to suppress enstrophy production at large scales and to reinforce the inverse cascade at the expense of the direct one. The coexistence of geometrical and dynamical flow confinement is a topic of recent investigation insofar restricted to idealized numerical studies in a triply-periodic domain \citep{deusebio14}. Flow confinement means horizontal anisotropy, namely the inhibition of flow evolution in the vertical, which splits the dynamics into two subranges of 3D isotropic and quasi-2D anisotropic turbulence, as summarized in Fig.~\ref{theory}.\\

Studies suggest the existence of two typical length scales, relative to both types of confinement, 
 that bound the region of flow anisotropisation, and therefore bound the inverse cascade. 
\begin{enumerate}
\item The first length scale is set by the aspect ratio $S = L_f/L_z$ between the typical length in the compacted direction $L_z$ and the scale of the forcing $L_f$ \citep{smith96}.  \cite{celani10} report that the energy transfer rate $\epsilon$ of the inverse cascade diminishes as the aspect ratio increases, and eventually vanishes for $S=1/2$. Even though this value is not expected to be
universal, it is useful to define a typical latitudinal index 
\begin{linenomath*}\begin{equation}
n_S = 2 \pi a/ H
\end{equation}\end{linenomath*}
\noindent corresponding to this threshold (here, $S = 1/2$ and $L_Z = H$). In the scale range $n<n_S$, geometrical confinement is suspected to lead to upscale energy transfer. 
\item A second length scale is provided by setting the non-dimensional Rossby number $Ro(n) = \epsilon^{1/3}/\Omega (\pi a/n)^{2/3}$ equal to one \citep{zeman94}. This defines a typical index 
\begin{linenomath*}\begin{equation}
n_{Ro} = \pi a (\Omega/\epsilon^{1/3})^{3/2}
\end{equation}\end{linenomath*}
\noindent that sets a scale range $n<n_{Ro}$ in which the flow is sensitive to the Coriolis effect, and develops barotropic modes (i.e dynamical confinement) \citep{smith99, deusebio14}. 
In this study, we estimate the Rossby number by integrating over all total indices $n$. Also, we will refer interchangeably to the Rossby index and the Rossby number to account for the strength of the dynamical confinement.
\end{enumerate}

Fig.~\ref{theory} shows that both the aspect ratio and Rossby indices ($n_S$ and $n_{Ro}$) are much smaller than the smallest scales resolved in our GCM (spectral analysis section~\ref{reference simulation} contains an estimate for $\epsilon$). Consequently, as is the case for all existing GCMs for gas giants, our simulations (with $1/2^{\circ}$ horizontal resolution) only partially reproduce the inverse cascade inertial range and do not reproduce the \textit{3D isotropic} domain at all. Note that because kinetic energy is created within the \textit{2D horizontal anisotropic} scale-range, an upscale energy cascade is expected.\\

\subsection{Spectral energy budget and zonal anisotropy}\label{Spectral energy budget}

The framework of zonostrophic turbulence extends Kraichnan's classical definition of the energy transfer range \citep{kraichnan71} to 2D turbulent flows with a $\beta$-effect \citep{galperin10}. The inverse energy cascade is zonally anisotropized,  leading to zonal and residual spectra in the following universal form \citep{sukoriansky02},
\begin{linenomath*}\begin{subequations}\label{KES}
\begin{align}
        E_Z(n)&=C_Z\beta^2 n^{-5}, \hspace{0.6cm} C_Z \approx 0.5, \label{KESa}\\
        E_R(n)&=C_R \epsilon^{2/3} n^{-5/3}, \hspace{0.6cm} C_R \approx 5-6.\label{KESb}
\end{align}
\end{subequations}\end{linenomath*}
The residual spectrum (black dashed line in Fig.~\ref{theory}) closely follows the classical Kolmogorov-Batchelor-Kraichnan (KBK) theory of 2-D isotropic turbulence, while the zonal spectrum (red dashed line in Fig.~\ref{theory}) clearly diverges from the non-axisymmetric modes with its steeper $-5$ slope.  The intersection of the zonal and residual spectra defines the transitional index 
\begin{linenomath*}\begin{equation}
n_{\beta} = (C_Z/C_R)^{3/10} (\beta^3/\epsilon)^{1/5}
\end{equation}\end{linenomath*}
\noindent at which the turbulent and Rossby wave frequencies are equal.  This length scale compares two processes that are competing in the system: the flow capacity to transfer energy uspscale through the value of $\epsilon$ (which depends on how flow bi-dimensionalization is efficiently set) versus its ability to channel energy in the zonal direction through parameter $\beta$ (here, the parameter $\beta$ is estimated at mid-latitude $\varphi = 45^{\circ}$). 
In simulations assuming 2D barotropic flows, \cite{sukoriansky08} and \cite{galperin10} find 
$(C_Z/C_R)^{3/10} \approx 0.5$, while \cite{galperin14} and \cite{cabanes17,cabanes18} find $\approx 0.75$ using observed Jupiter wind fields and laboratory experiments, respectively.\\

In classical isotropic 2D-turbulence on a non-rotating sphere, the inverse energy cascade ranges between the injection scale $n_{i}$ (here $\approx n_D$), at which the flow is energized, and the large-scale friction $n_{fr}$, at which energy is dissipated. In the case of a rotating sphere, the additional $\beta$-effect splits energy transfer into two subranges $n_{fr} \rev{<} n \rev{<} n_{\beta}$ and $n_{\beta} \rev{<} n \rev{<} n_i$ \citep{huang01,galperin10}. The former corresponds to the zonostrophic inertial range where the axisymmetric flow dominates and forms a system of multiple jets \citep{sukoriansky02,sukoriansky08,galperin10}. We call the latter  the residual-dominated inertial range, where the $\beta$-effect is weaker and the flow is dominated by the residual spectrum. These two subranges will be of crucial importance in the following study, as they are both fully resolved by the model. In flows with a strong $\beta$-effect, it is well accepted that the Rhines total index 
\begin{linenomath*}\begin{equation}
n_R = a(\beta/2U)^{1/2}
\end{equation}\end{linenomath*}
\noindent is a good approximation of the frictional scale at which the upscale cascade is arrested \citep{sukoriansky07}. Rhines' index also gives an estimate of the jets' size, with $U$ the root-mean-square (RMS) of the total velocity. The extension of the zonostrophic inertial range, and correspondingly the strength of the zonostrophic regime, can be estimated by the zonostrophy index $R_{\beta} = n_{\beta}/n_R$. It is generally accepted that zonostrophic turbulence emerges for $R_{\beta} \geq 2.5$ \citep{galperin06}.\\

Finally, there is an ultimate length scale at which $\beta$ becomes sufficiently weak that the flow reaches the domain of \textit{2D horizontal anisotropy} where the zonal and residual energy spectra are undifferentiated. This transition is observed in the modal spectrum, which is defined as 
\begin{linenomath*}\begin{equation}
	\mathscr{E}(n,m) = E_R(n)/(2n+1) = (1/2) \epsilon^{2/3} n^{-8/3}\label{eq:modalspectrum}
\end{equation}\end{linenomath*}
and is the green dashed line in Fig.~\ref{theory}. Equating the modal spectrum to Eq.~\eqref{KESa}, we obtain an estimate for the zonal index:
\begin{linenomath*}\begin{equation}
	n_z = (C_A/C_R)^{3/10} (2n_{\beta}/\Delta m)^{3/7} n_{\beta}
\end{equation}\end{linenomath*}
at which the flow transits to non-zonostrophic turbulence (here $\Delta m=1$). Three primary spectral ranges are then identified as the total index $n$ increases: the \textit{zonal anisotropic}, the \textit{2D horizontal anisotropic}, and the \textit{3D isotropic} ranges. The existence of two successive flow anisotropisations is specific to 3D zonostrophic turbulence. Due to our resolution limit, only the \textit{zonal anisotropy} range is resolved in our simulations. Note, however, that the configuration depicted in Fig.~\ref{theory} is relative to our Saturn model, and might change using other planets' properties or by changing the internal model parameters.\\

All spectra and length scales presented in Fig.\ref{theory} are verifiable predictions of the zonostrophic regime if a sufficiently large zonostrophy index $R_{\beta}$ is achieved. Up to now, 2D-turbulent numerical studies as well as spectral analysis of Jupiter observations have reported good agreements in both spectral slope and magnitude \citep{del09saturn,galperin10,sukoriansky12,galperin14}. 
In the present study, we extend the statistical analysis of zonostrophic turbulence to a 3D-turbulent model. Contrary to previous 2D barotropic studies, flow confinement in our GCM is not assumed but results from the shallow-rotating conditions that enable inverse energy cascade. Hence, geometrical $n_S$ and dynamical $n_{Ro}$ indices become two additional relevant scales of the system. The index $n_S$ is unchanged in all our simulations for a fixed atmospheric layer depth H. Thus, varying the rotation rate defines simulations with distinct Rossby numbers, which sets the strength of the upscale energy transfer.    

\section{Statistical flow analysis }\label{Statistical analysis}

\subsection{Kinetic energy spectra}
To compute the spectral analysis of the horizontal velocity fields $\mathbf{u}$ (i.e. a wind map at a fixed vertical position) it is appropriate to decompose the vector field in terms of two scalar functions. A decomposition into rotational (non-divergent) $\mathbf{v}$ and divergent (irrotational) $\mathbf{w}$ velocity components can be performed via the Helmholtz expression
$ \mathbf{u} = \nabla_h \times \psi \mathbf{e_z} + \nabla_h \phi  = \mathbf{v} + \mathbf{w}$, where scalar fields $\psi$ and $\phi$ are the horizontal streamfunction and velocity potential, respectively.
Using this decomposition, one obtains the vorticity $ \xi$ and the divergence $ \delta $ fields,
\begin{linenomath*}\begin{subequations}
\begin{align}
        \xi&=rot_h(\mathbf{u}) = \nabla_h^2 \psi\\
        \delta&=div_h(\mathbf{u}) = \nabla_h^2 \phi.
\end{align}
\end{subequations}\end{linenomath*}
By definition, the scalar product of two horizontal vector fields $\mathbf{a}$ and $\mathbf{b}$ can be expressed in the spectral domain as,
\begin{linenomath*}\begin{equation}\label{scalarproduct}
(\mathbf{a},\mathbf{b})_{m,n} = \frac{a^2}{n(n+1)} \mathbb{R} \left\lbrace rot_h(\mathbf{a})_{m,n} rot_h^{\ast}(\mathbf{b})_{m,n} + div_h(\mathbf{a})_{m,n} div_h^{\ast}(\mathbf{a})_{m,n} \right\rbrace
\end{equation}\end{linenomath*}
where $\ast$ is the complex conjugate and $\mathbb{R}\lbrace \cdot \rbrace$ refers to the real part \citep[see][]{boer83,augier13}.
It is clear from expression~\eqref{scalarproduct} that the 2D kinetic energy spectum (i.e. spectral energy density per modes $m,n$) is
\begin{linenomath*}\begin{equation}\label{modal-spectra}
\mathscr{E}(n,m) = \frac{(\mathbf{u},\mathbf{u})_{m,n}}{2} = \frac{a^2}{2n(n+1)} \mathbb{R} \left\lbrace \xi_{m,n} \xi^{\ast}_{m,n} + \delta_{m,n} \delta^{\ast}_{m,n}\right\rbrace,
\end{equation}\end{linenomath*} 
where $\xi_{m,n}$ and $\delta_{m,n}$ are the spherical harmonic coefficients of the vorticity and divergence fields. Following \cite{sukoriansky02}, we define the 1D energy spectrum as $E(n) = \sum_{m=-n}^n \mathscr{E}(n,m)$, which can be decomposed into a sum of the zonal and residual components, $E(n) = E_Z (n) + E_R (n)$. The zonal spectrum corresponds to the axisymmetric energy ($m=0$):
\begin{linenomath*}\begin{equation}
	E_Z (n) = \mathscr{E}(n,0)
\end{equation}\end{linenomath*}
while the residual spectrum  is a sum over all non-axisymmetric modes ($m \neq 0$):
\begin{linenomath*}\begin{equation}
	E_R (n) = 2 \sum_{m=1}^n \mathscr{E}(n,m).
\end{equation}\end{linenomath*}
In the text below, we will refer to the modal spectra $\mathscr{E}(n,m)$ as it corresponds to kinetic energy for each pair of modes $(n,m)$. 

\subsection{Spectral fluxes}
In order to better characterize the non-linear dynamics of eddy-eddy interactions in Saturn atmosphere, we compute non-linear enstrophy and energy fluxes, which provide further insight into the enstrophy and energy redistribution among scales. Starting with the vorticity equation,
\begin{linenomath*}\begin{equation}\label{vorticityEquation}
 \frac{\partial\xi}{dt} = - (\mathbf{v} \cdot \nabla )\xi - D,
\end{equation}\end{linenomath*} 
where $\mathbf{v}$ is the rotational velocity, and $D$ is the vorticity tendency due to divergent flow and other vorticity sources and sinks. In spectral space, the enstrophy equation is
\begin{linenomath*}\begin{equation}\label{entrophyEquation}
 \frac{\partial G_n}{dt} = J_n  + D_n^G,
\end{equation}\end{linenomath*}
with enstrophy $G =\frac{1}{2}\xi^2 $, and divergent effects, sources, and sinks contained within $D_n^G$.
The enstrophy interaction term is given by the enstrophy transfer function \citep{burgess13,read18},
\begin{linenomath*}\begin{equation}
J_n = -\frac{1}{4} \sum_{m=-n}^n  \left[\xi^{\ast}_{m,n} \left\lbrace \mathbf{v} \cdot \nabla \xi \right\rbrace_{m,n} + \xi_{m,n} \left\lbrace \mathbf{v} \cdot \nabla \xi \right\rbrace^{\ast}_{m,n} \right],
\end{equation}\end{linenomath*}
and the associated energy transfer function is
\begin{linenomath*}\begin{equation}
I_n = \frac{a^2}{n(n+1)} J_n.
\end{equation}\end{linenomath*}
The enstrophy and energy fluxes can now be computed:
\begin{linenomath*}\begin{equation}\label{Tspectralfluxes}
\Pi^Z_{n+1} = - \sum_{l=1}^n J_l \hspace{0.5 cm} \text{and} \hspace{0.5 cm}  \Pi^E_{n+1} = - \sum_{l=1}^n I_l
\end{equation}\end{linenomath*}
respectively. The integral over all wavenumbers $N$ is zero by construction and should also be equal to zero at $n=0$ as the energy transfer functions represent conservative processes.\\

\section{$1/2^{\circ}$ Saturn reference simulation}\label{reference simulation}

\subsection{Time evolution of the kinetic energy}\label{Time evolution of the kinetic energy}

\begin{figure}[tb]
  \centering
    \includegraphics[width=0.7\textwidth]{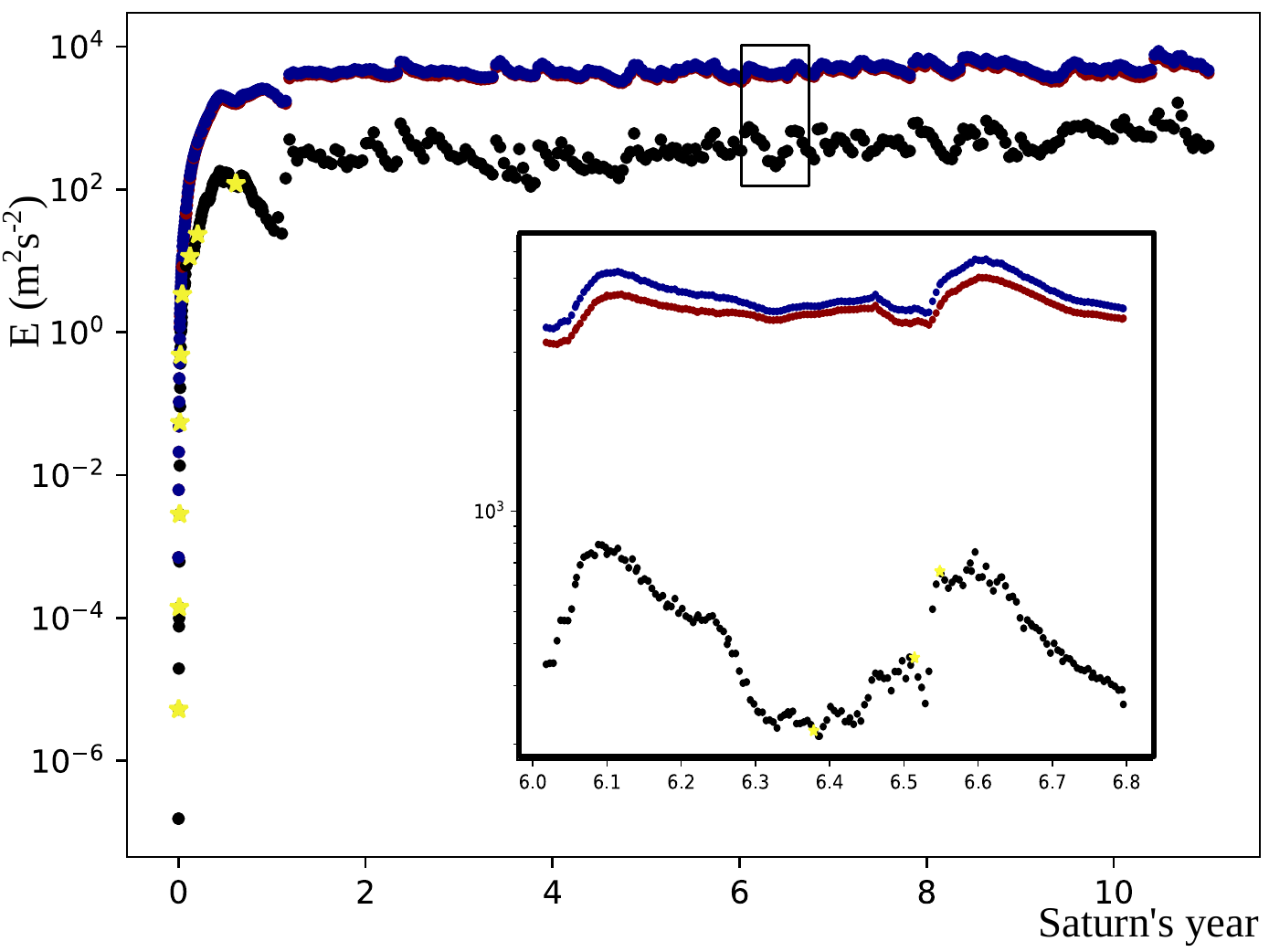}
  \caption{\rev{Time evolution of kinetic energy as a function of Saturn's simulated years:} time evolution of the zonal ($E_Z$ in red), residual ($E_R$ in black), and total energy ($E_T$ in blue), in Saturn years. Energy is in m$^2$\,s$^{-2}$. Yellow stars correspond to times $t \simeq 0.002$\,years (60\,days), 0.009 (220\,days), 0.012 (300\,days), 0.015 (380\,days), 0.02 (540\,days), 0.04, 0.1, 0.2, and 0.6 years. The black rectangle zooms in to the twelfth year, where the yellow stars are at times $t2=12.5$, $t3=12.78$, and $t4=12.8$ years.} \label{Energy-Tevo-IC}
\end{figure}

\citet{spiga2020} report on simulations covering 15 Saturn years using the Saturn DYNAMICO GCM. Wind fields are output every 20 Saturn days at 32 pressure levels onto $1/2^{\circ}$ latitude-longitude grid maps. 
The time evolution of the total energy  $E_T = E_Z + E_R$, as well as the zonal and residual energy, $E_Z= \sum_{n=0}^N E_Z(n)$ and $E_R = \sum_{n=0}^N E_R(n)$ respectively, are reported in Fig.~\ref{Energy-Tevo-IC}, with maximum total index $N = 360$ for $1/2^{\circ}$ horizontal resolution. The total energy strongly increases to reach a plateau after 1 Saturn simulated year. In the steady state, the zonal energy is an order of magnitude larger than the residual energy, and remains approximately constant over the 15 simulated years. The ratios of the time-average zonal and residual energy to the total energy over the two last simulated years are $\hat{E_Z}/\hat{E_T} \simeq 0.92$ and $\hat{E_R}/\hat{E_T} \simeq 0.08$, showing strong flow anisotropy in the zonal direction.  The energy profiles fluctuate around a mean value with two dominant frequencies corresponding to annual and semi-annual periods. A typical annual period is shown in the black square of Fig.~\ref{Energy-Tevo-IC}, which zooms into the twelfth simulated Saturn year. In what follows, we refer to these annual and semi-annual periods as seasonal energy cycles.

\subsection{Horizontal velocity and vorticity fields}\label{Horizontal velocity and vorticity fields}
\begin{figure}[tbp]
  \centering
    \includegraphics[width=\textwidth,clip,viewport=30 20 1090 800]{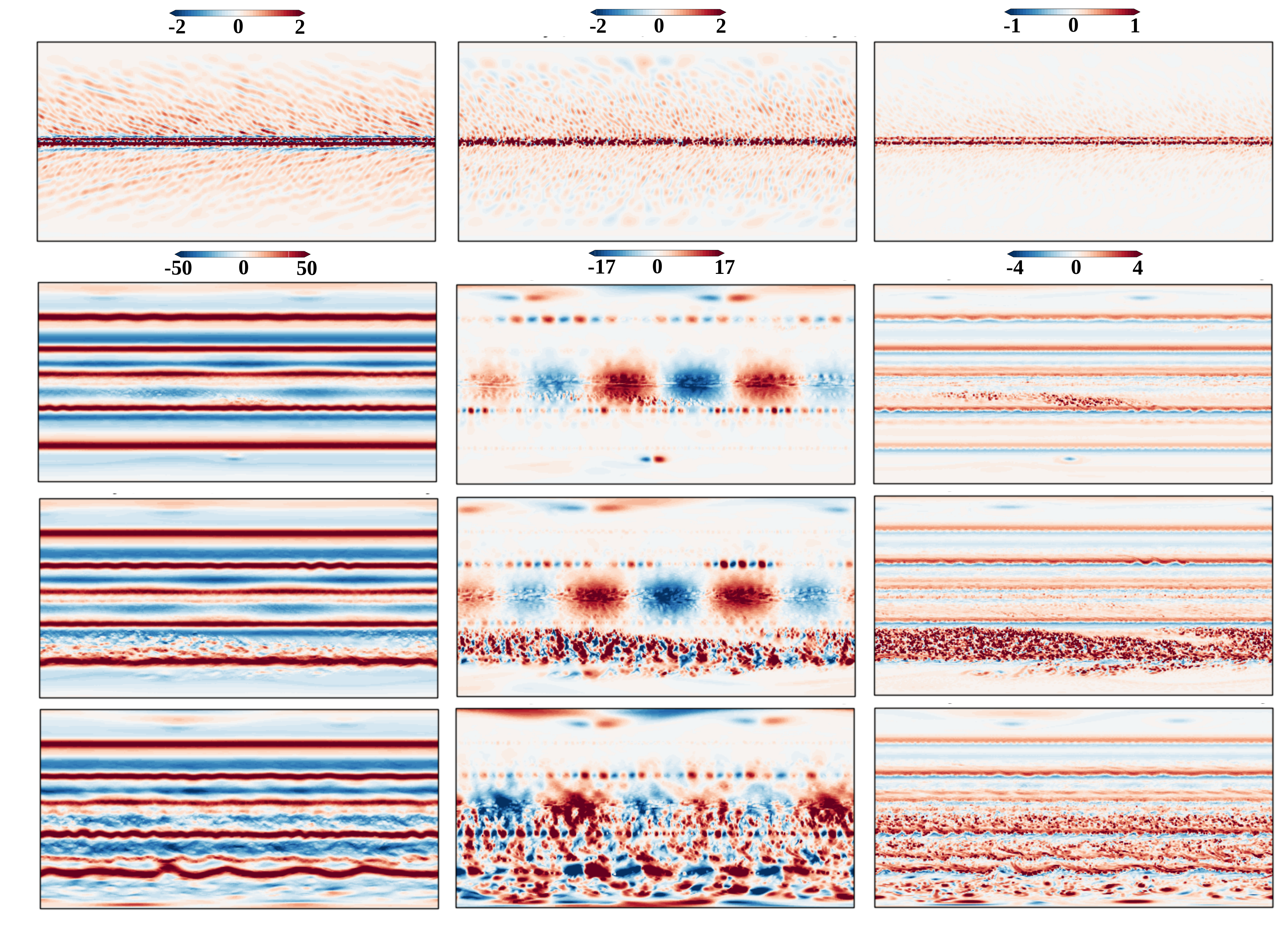}
    \caption{Instantaneous (latitude-longitude) \rev{velocity and vorticity} maps: zonal velocity (m\,s$^{-1}$) (left), meridional velocity (m\,s$^{-1}$) (middle) and vorticity fields (s$^{-1}$) (right), at altitude 94,28 km, at times $t1=0.04$, $t2=12.5$, $t3=12.78$, and $t4=12.8$  years. 
    These particular times are indicated with yellow stars in Fig.~\ref{Energy-Tevo-IC}.
    \label{refsimu-HorzSection}}
\end{figure}

Fig.~\ref{refsimu-HorzSection} shows instantaneous zonal and meridional velocity fields as well as the vorticity field at time step $t1=0.04$ Saturn year and 
once steady state is achieved. 
All wind fields are shown at 94.28 km above the 3 bar level (located on vertical cross-section Fig.~\ref{VerticalSection}), roughly corresponding to Saturn's visible tropospheric cloud deck. After time $t1=0.04$ Saturn years, baroclinic instabilities drive zonal and meridional motions with comparable magnitude, preferentially in the equatorial region.\\

In the steady state, the zonal velocity component largely dominates the flow, forming multiple axisymmetric jets with mid-latitude jets that reach 60--70\,m\,s$^{-1}$. The number of jets is consistent with the most recent observations (i.e. three per hemisphere) \citep{porco05,garcia10}. However, our model underestimates the jet magnitude by a factor 1.5--2. The prograde equatorial jet is only weakly super-rotating, and its intensity is approximately one order of magnitude smaller than observations. This means that jet acceleration by waves and eddies, albeit significant, remains underestimated with our model's current configuration \citep{spiga2020}. The three lower maps in Fig.~\ref{refsimu-HorzSection} correspond to a sequence of increasing residual energy in the annual cycle observed in Fig.~\ref{Energy-Tevo-IC}. We show times $t2=12.5$, $t3=12.78$, and $t4=12.8$ years (see the yellow stars Fig.~\ref{Energy-Tevo-IC}), which correspond to the minimum, intermediate, and maximum of residual energy in the zoomed in part of Fig.~\ref{Energy-Tevo-IC}, respectively. This sequence begins at $t2=12.5$ years with a pattern of steady jets, minimizing the non-axisymmetric perturbations. At $t3=12.78$ years, an instability grows from a jet and propagates throughout the entire southern hemisphere (referred to by \citet{spiga2020} as ``eddy bursts'' following the early work by \citet{Pane:93}). 
The system finally reaches a maximum of residual energy near $t4=12.8$ years, before it returns to a steady jet pattern with low residual energy. The same sequence occurs in the northern and southern hemispheres within the annual cycle. These energy cycles are due to the intermittent forcing that relates to the annual and seasonal variations of sunlight absorption in the atmosphere.

\subsection{Meridional cross-section of the velocity field}
\begin{figure}[tb]
  \centering
    \includegraphics[width=\textwidth]{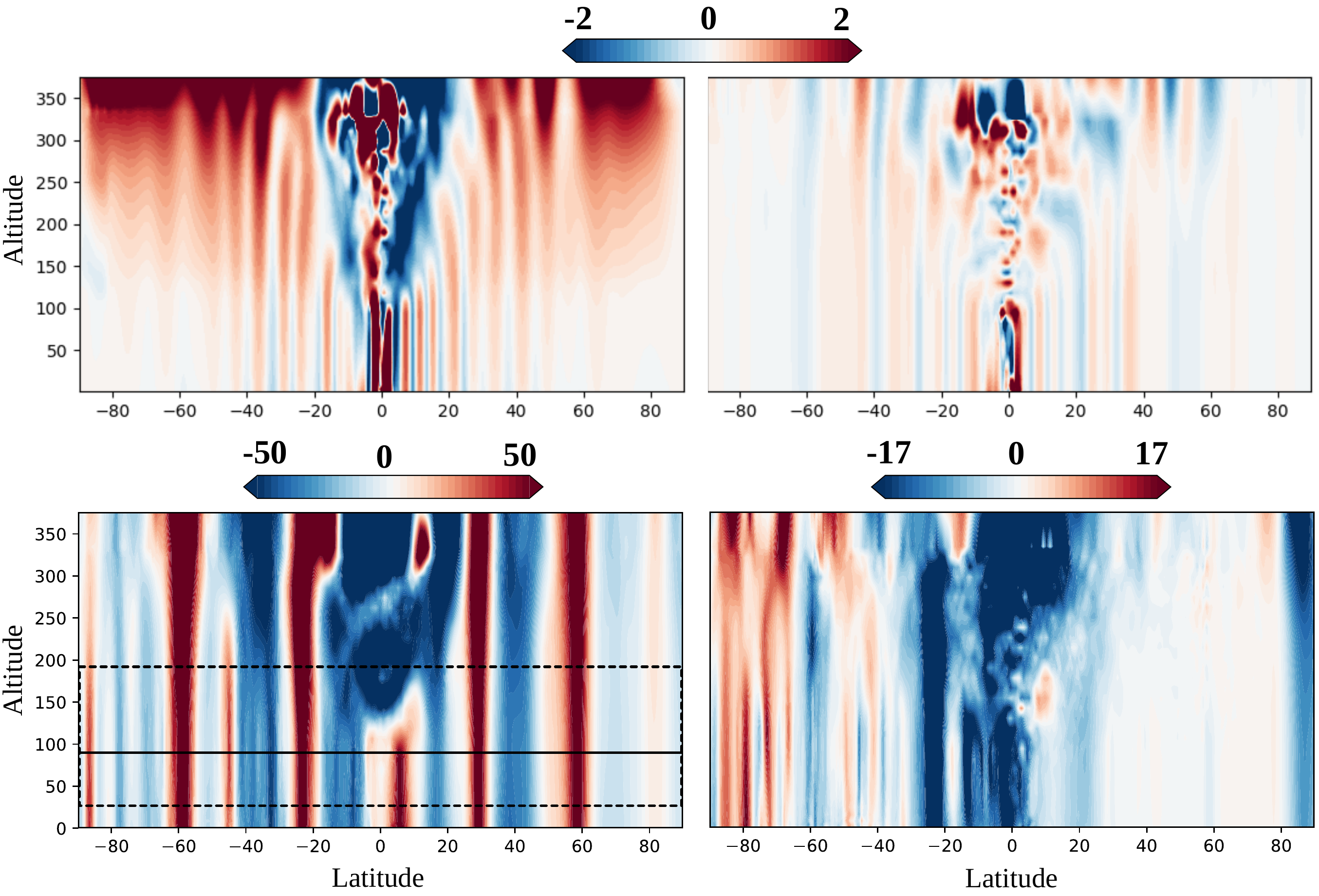}
    \caption{Instantaneous (altitude-latitude) \rev{meridional cross-section of velocity maps:} zonal (left) and meridional (right) velocities (m\,s$^{-1}$) at longitude $\lambda = 0^{\circ}$, times $t2=0.04$ (top) and $t4=12.5$ (bottom) Saturn years, as a function of latitude and altitude (in km, with 0 km at the 3-bar level, which is the bottom of the model). In the bottom left panel, the solid line corresponds to the level at which horizontal maps are shown. The black dashed square shows the vertical depth over which the statistical analyses are averaged.  \label{VerticalSection}}
\end{figure}

Figure~\ref{VerticalSection} shows meridional cross-sections through the zonal and meridional velocities at time $t1=0.04$ and $t4=12.5$ Saturn years. At the earlier time, baroclinic instabilities force zonal and meridional motions in the topmost layers, while the bottom velocity is close to zero. A strong velocity gradient prevails in the vertical and flow barotropization (i.e. vertical invariance) is only perceptible in the equatorial region. Nonetheless, the presence of a velocity gradient during the early simulated days provides evidence for vertical anisotropy in the forcing. Once the steady state is achieved at the later time shown, the flow is dynamically constrained to have a weak vertical gradient at all latitudes (i.e. it becomes strongly barotropic). 
The jets extend throughout the domain depth \rev{and form} a quasi-2D dynamical system in which the barotropic component of the flow accumulates most of the kinetic energy in Saturn's global circulation. As suggested by \cite{charney71}, flow barotropization can be related to an upscale energy cascade into the vertical flow component. Energy accumulation into the barotropic mode reaches its maximum in the troposphere and lower stratosphere (depicted as a dashed black square in Fig.~\ref{VerticalSection}) where the barotropic energy exceeds its baroclinic counterpart by a factor 7 while it decreases to a factor 3 in the upper stratosphere. This might relate to the continuous energetic forcing by baroclinic instabilities that occurs in the topmost stratospheric region. We also note that barotropization appears to be mainly active at high- and mid-latitude consistently with what have been observed by \cite{chemke15} for the Earth's atmosphere. Note that due to the shallow-atmosphere approximation, where the latitudinal projection of the rotation spin axis is neglected, these barotropic columns align with the spherical radius. By implementing all components of the Coriolis force in the model equations, these columns would likely tilt in latitude. \rev{Such flow barotropization is expected to extend down to depths of thousands of kilometres beneath the tropospheric cloud level, probably into the region of magnetic dissipation at a depth of $\sim 3,000$ km for Jupiter \citep{kaspi18} and $\sim 9,000$ km for Saturn \citep{guillot18,Gala:19}. Here, our Saturn reference simulation limits us to a thin atmosphere of $\sim 360$ km in the vertical, which remains a key intrinsic limitation of shallow GCMs. }
 
\subsection{Spectral analysis of the wind field}\label{Spectral analysis of the wind field}

\begin{figure}[tb]
  \centering
    \includegraphics[width=0.9\textwidth]{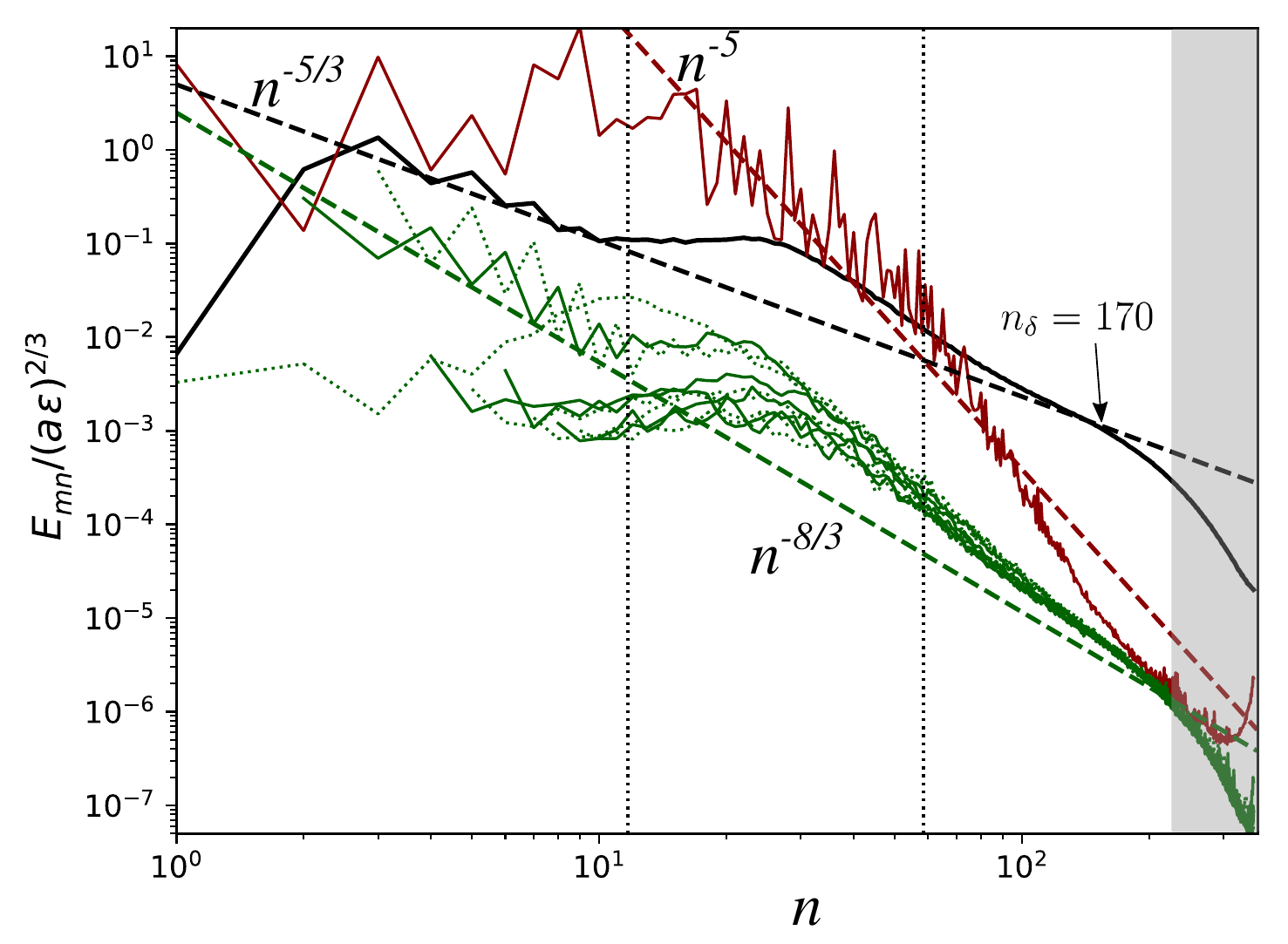}
  \caption{\rev{Spatial kinetic energy spectra as a function of latitudinal index:} vertical grey lines are the Rhines scale $n_R$ (left) and the transitional scale $n_{\beta}$ (right). The grey area corresponds to the Rossby deformation radii $n_D$, showing typical scales of the baroclinic instabilities. Normalized zonal and residual energy spectra $E_Z(n)/(\epsilon a)^{2/3}$ (red) and $E_R(n)/(\epsilon a)^{2/3}$ (black) as well as modal spectra for $m=1$ to $10$, $\mathscr{E}(n,m)/(\epsilon a)^{2/3}$ (green). Dashed lines represent theoretical predictions from expressions: Eqs~\eqref{KESa} in red, \eqref{KESb} in black and Eq~\eqref{eq:modalspectrum} in green. \label{Sprectra-Complet}}
\end{figure}

At different vertical levels we computed the horizontal velocity spherical harmonic decomposition, following Eq.~\eqref{modal-spectra}. Figure~\ref{Sprectra-Complet} summarizes the kinetic energy spectra in the steady state for our Saturn reference simulation, showing the zonal spectrum $E_Z(n)$, the residual spectrum $E_R(n)$, and the modal spectrum $\mathscr{E}(n,m)$ (for modes $m=1$ to 10 only). All spectra are averaged in time over the two last simulated years, and over several vertical levels sampling both the troposphere and the lower stratosphere, i.e. where energy strongly accumulates into the barotropic mode (see black dashed square in Fig.~\ref{VerticalSection}). \\

The zonal spectrum follows the -5 slope and magnitude predicted by Eq.~\eqref{KESa}. At large scales, the zonal spectrum exceeds the residual spectrum by about two orders of magnitude, providing evidence for the strong zonal anisotropy inherent to zonostrophic turbulence. Although $m=1$, 2, and 3 are more energetic and bumpy than other modes, the modal spectra preserve a nearly $n^{-8/3}$ slope over a wide range of $n$.  
Likewise, the global trend in the residual spectrum over the full range of $n$ fits the $n^{-5/3}$ slope expected from the KBK theory of 2D turbulence, and makes it possible to estimate the inverse energy cascade rate $\epsilon$. The bump in the residual and modal spectra at moderate $n$, which makes these spectra differ at medium scales from the $n^{-5/3}$ and $n^{-8/3}$ slopes respectively, will be discussed below. Using Eq.~\eqref{KESb} and assuming $C_R = 6$, we estimate $\epsilon \approx 2 \times 10^{-6}$ m$^2$\,s$^{-3}$. \rev{Considering that the total kinetic energy growth during model spinup will be approximately linear, i.e. $E_T(t) = \epsilon t$, and that the spinup period lasts about one Saturn simulated year or $10^9$ s (Fig.~\ref{Energy-Tevo-IC}), the equilibrium total kinetic energy of about 2000 m$^2$ s$^{-2}$ in Fig.~\ref{Energy-Tevo-IC} corresponds very well to the total energy injected during the first year of the simulation.} With $\beta = 4 \times 10^{-12}$ m$^{-1}$\,s$^{-1}$ and $C_Z = 0.5$, we compute the transitional index $n_{\beta} \simeq 58$, the zonal index $n_Z \simeq 449$, and the Rossby index $n_{Ro} \simeq 2.7 \times 10^{5}$ and non-dimensional number $Ro = 4 \times 10^{-3}$ for our Saturn reference simulation. We estimate the RMS velocity $U \simeq 55$ m\,s$^{-1}$ using the total energy in the steady state $\hat{E_T}$, leading to a value $n_R \simeq 11$ for the Rhines index. In Fig.~\ref{Sprectra-Complet}, the Rhines and transitional indices border the zonostrophic inertial range characterized by its zonostrophy index $R_{\beta} = 5.3$, which significantly exceeds the zonostrophic threshold $R_{\beta} \geq 2.5$ stated in Section~\ref{Spectral energy budget}. These statistics are summaried in Table~\ref{tab:energydiags}. The estimated value of $n_Z$ exceeds the largest index $N=360$ resolved in our $1/2^{\circ}$ Saturn reference simulation, and so does not appear in Fig.~\ref{Sprectra-Complet}.  \\

By approximating the zonal and residual spectra with their predictions, Eqs.~\eqref{KESa} and \eqref{KESb}, assuming straight horizontal lines for $n\leq n_{fr}$, and by integrating over indices $n$, one finds analytical expressions for $E_Z$ and $E_R$ \citep{sukoriansky02,galperin14}. This leads, in the framework of zonostrophic turbulence, to the definition of the anisotropy index
\begin{linenomath*}\begin{equation}
\gamma  = E_Z/E_R\approx (3/4)(10C_Z)^{-5/6}R_{\beta}^{10/3}.
\end{equation}\end{linenomath*}
The associated ratios $E_Z/E_T = \gamma/(1+\gamma)$ and $\hat{E_R}/\hat{E_T} = 1/(1+\gamma)$ give $E_Z/E_T \simeq 0.98$ and $E_R/E_T \simeq 0.02$ with $R_{\beta} = 5.3$. This sightly overestimates the time-averaged ratios obtained in Section~\ref{Time evolution of the kinetic energy} (i.e. $E_Z/E_T \simeq 0.92$ and $E_R/E_T \simeq 0.8$) from spectral analysis of the simulated flow. Both estimates nevertheless point towards significant flow anisotropy.\\

\begin{table}
\centering
 \begin{tabular}{c c c c c c } 
 \toprule
   & $\epsilon$ (m$^2$\,s$^{-3}$)  & $n_Z$& $n_{\beta}$ & $n_R$ & $R_{\beta}$ \\ [0.5ex] 
 \midrule
 Saturn reference simulation & $2 \times 10^{-6}$ & 449 & 58 & 11 & 5.3 \\
 \cite{choi11} & $5 \times 10^{-6}$& 550 &  67  &  10.9 & 6.1\\
 \cite{galperin14} & $10^{-5}$ &  430 & 57   & 10.9 & 5.2\\ 
 \bottomrule
\end{tabular}
  \caption{Summary of diagnostic quantities that facilitates comparison between our Saturn reference simulation (SRS) and previous studies of Jupiter's tropospheric dynamics using cloud tracking from Cassini images. } \label{tab:energydiags}
\end{table}

The statistical analysis of our Saturn simulation is consistent with all basic properties of the theoretical scaling introduced in Section~\ref{Scaling in 3D zonostrophic turbulence} and summarised in Fig.~\ref{theory}. 
The non-dimensional scaling used in this study allows \rev{for} direct comparison with other planetary atmospheres in which zonostrophic turbulence develops.  \rev{Statistical analyses of Jupiter's atmospheric circulation at cloud level have been previously described} by \cite{choi11} and \cite{galperin14}  \rev{using} Jupiter images from the Cassini mission. 
\rev{\cite{galperin14} calculated non-dimensional scalings similar to those used in the present study based on both
their own wind field data set and \cite{choi11}'s.}
The \rev{resulting} non-dimensional quantities \rev{are} reported in Table~\ref{tab:energydiags} \rev{and} show that the Jovian troposphere and our Saturn numerical simulation share analogous dynamics with similar zonostrophy indices and spectral anisotropy. However, specific features of Saturn's dynamics appear when looking at the details of the energy spectra. In our simulation, the modal spectra $\mathscr{E}(n,m)$ show that modes $m= 2$ and $3$ make a dominant contribution to the residual spectrum, suggesting strongly energetic waves that propagate in the zonal direction. \citet{spiga2020} identified indeed wavenumber-2 and wavenumber-3 Rossby waves and Rossby-gravity waves, with possible relation to wave patterns observed by either Voyager \citep{achterberg96} or Cassini \citep{guerlet18}.
However, as of now, a complete statistical analysis using Saturn direct observations is yet to be performed on the most recent images. Consequently, prominent modes and spectral energy bumps in our Saturn reference simulation remain to be confirmed by observations.\\

Our simulation captures the global statistical picture of the gas giants and specifically Saturn's dynamics in some respects, despite the \textit{ad hoc} conditions enforced in our GCM to compensate for unknown frictional effects and unresolved small scale dissipation. Hypeviscosity is responsible for the sudden energy drops where all spectra significantly deviate from predictions. It qualitatively defines a typical range of indices, $n_{\delta}\gtrsim 170$, for which hypeviscosity effects are predominant. 
Frictional effects are less evident but a likely crucial to set the large-scale dynamical regime in which a wealth of turbulent waves and eddies develop. Our model also suffers intrinsic limitations such as limited vertical resolution and the absence of moist convection, which are both likely to impact wave propagation and global atmospheric circulation. \\

\subsection{Spectral fluxes and time evolution}\label{Spectral fluxes and time evolution}

\begin{figure}[p]
  \centering
    \includegraphics[width=\textwidth]{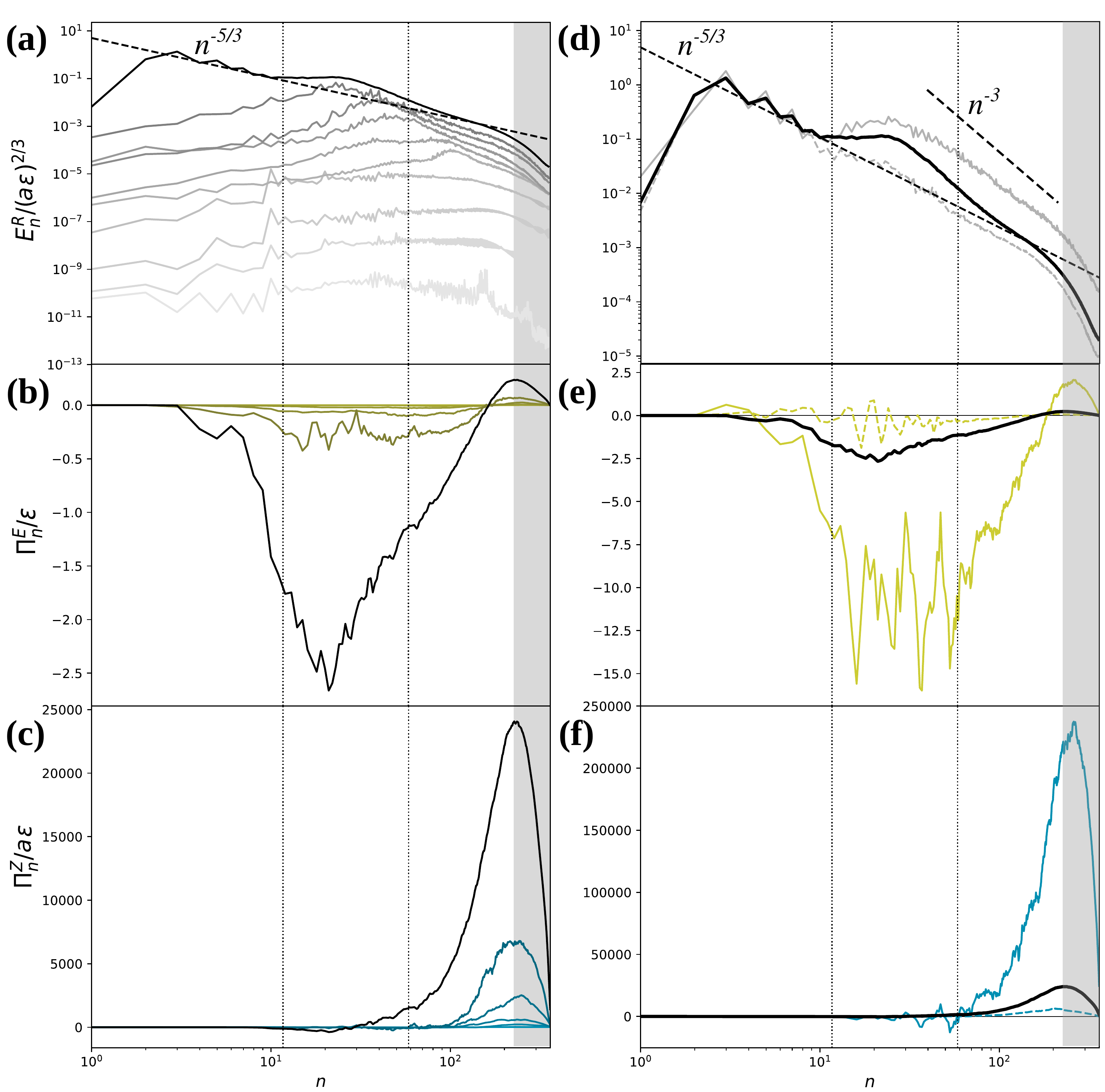}
      \caption{\rev{Time evolution of spatial kinetic energy spectra as well as energy and enstrophy fluxes as a function of latitudinal index:} vertical grey lines are the Rhines scale $n_R$ (left) and the transitional scale $n_{\beta}$ (right). The grey area corresponds to the Rossby deformation radii $n_D$, showing typical scales of the baroclinic instabilities. Normalized (top) residual kinetic energy spectra $E_R(n)/(\epsilon a)^{2/3}$, (middle) spectral fluxes of energy $\Pi^E/\epsilon$ and (bottom) spectral fluxes of enstrophy $\Pi^Z/(\epsilon^{-1} a^2)$. In all panels, spectral quantities in the steady state are averaged in time and shown as solid black lines. (Left) Transients of the instantaneous spectral quantities from the start of the simulation (colour gradient) to the steady state (black line). The colour gradient from light to dark lines corresponds to times $t \simeq 0.002 $ (60\,days), 0.009 (220\,days), 0.012 (300\,days), 0.015 (380\,days), 0.02 (540\,days), 0.04, 0.1, 0.2, and 0.6 years. (Right) Instantaneous spectral quantities near the minimum of the energy cycle at $t2=12.5$ years (coloured dashed lines) and the maximum of the energy cycle at $t4=12.8$ years (coloured solid lines) of the annual sequence shown in Fig.~\ref{Energy-Tevo-IC}. \label{Spectra-Transitory}}
\end{figure}

We present in Fig.~\ref{Spectra-Transitory} transient dynamics showing instantaneous kinetic energy spectra as well as energy and enstrophy spectral fluxes. Time evolution of the spectral quantities are shown in Fig.~\ref{Spectra-Transitory}(a-c) from an atmosphere at rest to the steady state, and in (d-f) we show the annual oscillation in energy diagnostics, which can be interpreted as the standard deviation of the mean state. These instantaneous spectra correspond to the time steps marked with yellow stars in Fig.~\ref{Energy-Tevo-IC}.  
In what follows, the terms ``steady state'' and ``mean state'' are used synonymously to refer to time-averaged spectral quantities using the averaging procedure described in Sect.~\ref{Spectral analysis of the wind field} (and Fig.~\ref{Sprectra-Complet}). This section provides more details on non-linear interactions and general energetic pathways in Saturn's atmospheric circulation throughout our numerical simulation. More specifically, this approach allows us to identify the scale at which kinetic energy is created in the system, how it transits, and where it ends up.

Fig.~\ref{Spectra-Transitory}(a) shows the time evolution of the residual kinetic energy spectra from 60 simulated days to half a year before it reaches the steady state. Time steps are located in time on the time evolution profiles of kinetic energy in Fig.~\ref{Energy-Tevo-IC}. At the first time step, the spectrum is weakly energetic and approximately flat over all scales. When the flow is energized by solar absorption in the stratosphere, the spectra show that kinetic energy is predominantly increased at small scales. 
As mentioned above, the most relevant length scales for energy creation are the baroclinic Rossby deformation radii. However, hypeviscosity presumably damps kinetic energy in the scale range that co-incides with the typical length scales of baroclinic eddies. 
This damping is likely to impede the formation of a sharp energy peak near $n_D$ and certainly reduce the global energy budget at all scales. Such a condition at small scales suggests that our simulation is eddy-permitting but not fully eddy-resolving. Under confinement effects\rev{, i.e. the coupled action of geometrical and dynamical confinement,} 
the energy spectra develop a $n^{-5/3}$ slope with a bumpy front that propagates in time toward large scales, up to $n\simeq 2$. This upscale energy transfer is not arrested at the Rhines scale, and extends over a large spectral domain, i.e. $2 \lesssim n \lesssim 170$, for which the  $n^{-5/3}$ slope fits the residual spectrum very well in the steady state.\\

Figure~\ref{Spectra-Transitory}(b,c) shows snapshots and the steady state of the spectral fluxes at the same time steps. Positive spectral flux corresponds to energy transfer from small to large indices $n$, which is a downscale energy cascade, and conversely negative values correspond to an upscale energy cascade. The general trend in the steady state is clear: energy flux is negative from $n_{\delta} = 170$ down to $n=2$, showing a wide inertial range with upscale energy transfer. 
This is consistent with the $n^{-5/3}$ spectral signature of residual kinetic energy shown in Fig.~\ref{Spectra-Transitory}(a). At scales $n_{\delta} \rev{\gtrsim} 170$, the positive energy flux suggests a downscale energy cascade over a range that perfectly matches the spectral domain where hyperviscosity directly affects the kinetic energy spectra. The index $n_{\delta}=170$ at which energy flux changes sign appears to be robust at all times during spin-up, and occurs at a scale slightly larger than the first Rossby deformation radius with typical index $n_D = 225$. hypeviscosity possibly distorts energy creation, shifting to larger scale the onset of the upscale cascade. When the flow satisfies the conditions for quasi-geostrophy, theoretical studies \citep[e.g.][]{charney71,salmon80} also suggest the existence of an downscale enstrophy cascade. This trend seems to be consistent with our Saturn reference simulation: in Fig.~\ref{Spectra-Transitory}(c), the enstrophy flux is predominantly positive and increases strongly with $n$ up to the resolution limit. At scales larger than the transitional index $n_{\beta}$, the enstrophy flux becomes weakly negative, showing a tendency for an upscale enstrophy cascade. Such a feature has also been observed for Jupiter by \cite{young17}, in which the upscale enstrophy transfer induces a weak peak in enstrophy near the Rhines scale.\\%

In Fig.~\ref{Spectra-Transitory}(d-f), we bracket the mean state with two instantaneous plots of spectral quantities close to the minimum (dashed lines) and the maximum (thin solid lines) of residual energy in the annual cycle described in Sect.~\ref{Time evolution of the kinetic energy} and presented in Fig.~\ref{Energy-Tevo-IC}. Spectral instantaneous diagnostics are computed at times $t2=12.5$ and $t4=12.8$ years. The spectra in Fig.~\ref{Spectra-Transitory}(d) reveal interesting energy exchanges once the steady state is achieved: the Rhines scale splits the dynamics between long-lived modes, which control the large scales for indices $n<n_R$, and transitional modes, which develop and break up within the zonostrophic inertial range (i.e. $n_R<n<n_{\beta}$) during the annual energy cycle.

On the one hand, in the large scale range $n<n_R$, the long-lived structures are dominated by two highly energetic zonal modes, $m=2$ and $m=3$, whose amplitude is insensitive to the annual variations. Thus, we refer to those as long-lived modes: non-stationary $m=2$ and $m=3$ inertial waves which are time-independent when looking at spatial spectra $E_R$.  This is illustrated in Fig.~\ref{Spectra-Transitory}(d), where instantaneous spectra are similar to the spectrum of the mean state in the range $n<n_R$. This is also observed in physical space when looking at meridional velocity maps (Fig.~\ref{refsimu-HorzSection}), for which the $m=3$ mode is persistent over the whole energy sequence shown \citep[the $m=2$ mode was shown in][Fig.~13]{spiga2020}. 

\begin{figure}
  \centering
    \includegraphics[width=\textwidth]{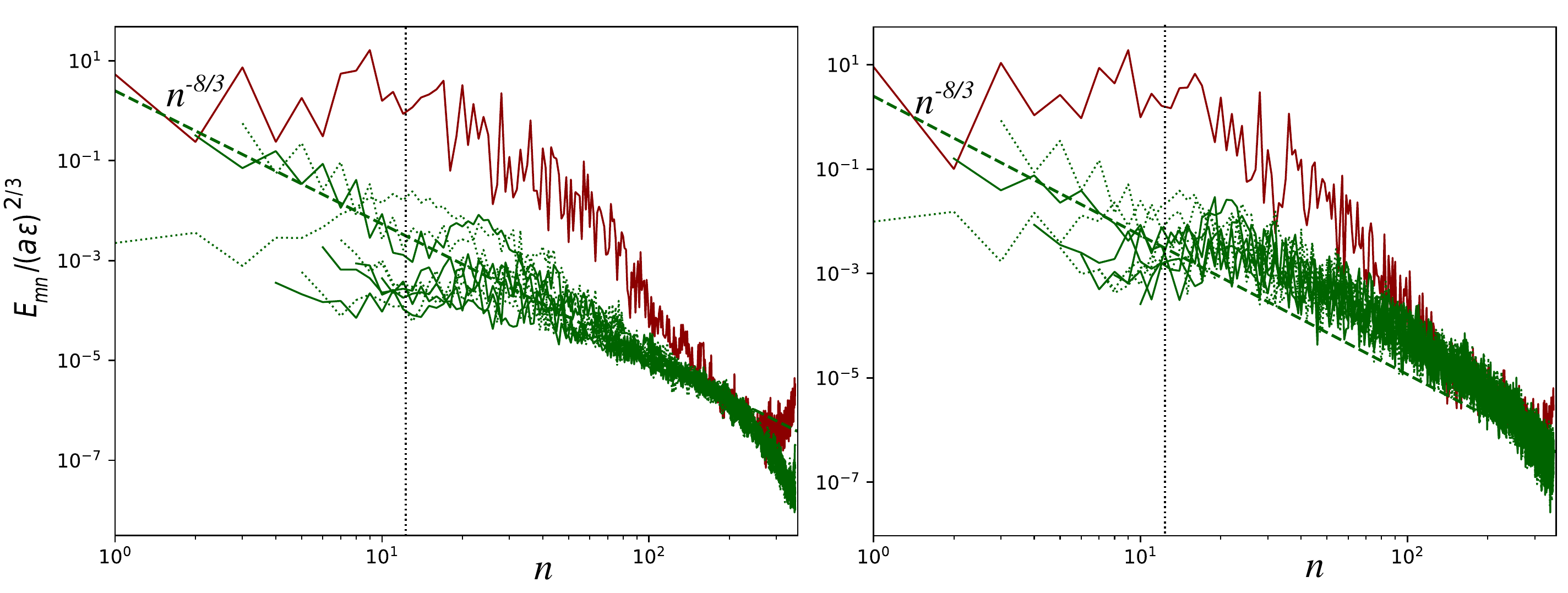}
      \caption{\rev{Instantaneous modal kinetic energy spectra as a function of latitudinal index:} normalized zonal energy spectra $E_Z(n)/(\epsilon a)^{2/3}$ (red) and modal spectra for $m=1$ to $10$, $\mathscr{E}(n,m)/(\epsilon a)^{2/3}$ (green). Left panel corresponds to the minimum of energy  at $t2=12.5$ years and right panel to the maximum of energy  at $t4=12.8$ years, in the annual energy cycle shown in Fig.~\ref{Energy-Tevo-IC}. Dashed lines represent theoretical predictions of the modal spectra from Eq~\eqref{eq:modalspectrum} and Rhines index $n_r$ (vertical dotted lines).
\label{Transient-Modalspectra}}
\end{figure}

On the other hand, in the zonostrophic range, the residual spectrum oscillates around the mean state showing a strongly energetic phase with a nearly $n^{-3}$  slope, and a phase of low energy matching the KBK $n^{-5/3}$ slope.   \citep{chertkov07}. \rev{The $n^{-3}$ spectral signature and the associated bumpy residual spectrum might be related to the instability of the zonal jets, which are barotropically unstable according to the Rayleigh-Kuo criterion (see \cite{ingersoll82} for the general perspective, and \cite{spiga2020} for the SRS specifically). To support this,} the modal spectra in Fig.~\ref{Transient-Modalspectra}, show that \rev{the strongly energetic phase,}  is \rev{related to an enrichment of}  a large variety of zonal modes that result from jet destabilization \rev{ due to barotropic instability}. This intermittent enrichment by zonal modes of the energy spectra is dominantly observed at scales that are \rev{close to but} smaller than the Rhines scale, making the spectral slope at $n>n_R$ much steeper. We note that the zonal spectrum remains unchanged over time, and so are the jets.\\

Similarly, snapshots of the spectral energy flux over the annual cycle (Fig.~\ref{Spectra-Transitory}e) show that the extended domain of the upscale cascade in Fig.~\ref{Spectra-Transitory}(b) results from the averaging procedure over time, 
and does not account for instantaneous dynamics.
Instantaneous diagnostics of the energy flux show strong perturbations and sign changes, particularly in the range $n<n_{\beta}$, which suggests a superimposition of direct and inverse energy cascades. \rev{Again, when jets become barotropically unstable they break into turbulence and feed a direct  cascade of energy that appears to be of a comparable strength to the inverse cascade in the low energetic phase.}
Nonetheless, when the residual energy is at a maximum, the upscale cascade dominates except for a small scale range, i.e. $n \lesssim n_R$, where energy converges both from large and small scales as energy flux changes sign at the largest scales. Such a convergence of kinetic energy at the jet scale has been reported by \cite{young17}, who computed spectral fluxes from snapshots of Jupiter's dynamics (covering 3--4 planetary rotation periods only). We also note that, within the zonostrophic inertial range, local perturbations of the energy flux (with a serrated shape) are associated with similar perturbations of the enstrophy flux, which departs from the classical picture of 2D turbulence (see Fig.~\ref{Spectra-Transitory} e-f). This erratic enstrophy production at scales larger than the transitional scale $n_{\beta}$, only visible in instantaneous fluxes, is suggestive of vorticity filamentation that develops in the zonostrophic inertial range of 3D-turbulent atmospheric layers.\\

In this section we have explored the dynamics of a 3D-turbulent atmospheric layer that reproduces the statistical properties of zonostrophic turbulence. Under geometric and dynamical confinements, time-averaged spectral quantities are in good agreement with the energy-enstrophy double cascade scenario. This double cascade found in our 3D simulations may be reminiscent of
the double cascade described in 2D flows  by \cite{Kraichnan67} theory. The statistical analysis of our Saturn reference simulation resembles in several respects the spectral quantities reported from Jupiter images \citep{young17} and other GCMs \citep{schneider09,read18}. However, a novelty of our statistical analysis is that by looking at the instantaneous spectral quantities we reveal peculiar relevance of the Rhines and transitional scales in the 3D-turbulent zonostrophic regime. The Rhines scale seems to act as an energetic barrier that splits the dynamics between long-lived modes, i.e. dominantly $m=2$ and $m=3$, and intermittent modes, which strongly depend on the seasonal energy cycles. Instantaneous quantities also reveal the superimposition of the inverse and direct cascades of both energy and enstrophy, in the extended inertial range $n<n_{\beta}$. These complex dynamics are likely related to the formation of inertial modes, which develop and undergo vorticity filamentation in 3D-turbulent atmospheric layers driven by an intermittent forcing that reflects the seasonal energy cycles. In the next section we explore the role played by these seasonal energy cycles in the global flow statistics by suppressing radiative transfer computations in the GCM and implementing instead a continuous idealized Taylor-Green forcing that is time-independent.

\section{Idealized simulations}\label{Idealized simulations}

In order to explore the regime of Saturn macro-turbulence and other open questions about the nature of the atmospheric circulation of giant planets, we run a set of four idealized numerical simulations with varying rotation rate and a well-controlled energy forcing.

We use an idealized version of our Global Climate Model based on DYNAMICO in which atmospheric flow in a Saturn-like atmosphere is driven by a Taylor-Green forcing \citep{alexakis15} and flow baroclinicity is suppressed \citep[this is obtained by switching off the physical parameterizations described in][notably radiative transfer]{guerlet14}. Kinetic energy is injected every 32 dynamical time steps at all vertical levels at a prescribed horizontal wavenumber $k_i$. The forcing is given by 
\begin{linenomath*}\begin{equation}
\mathbf{f}=\left(
  \begin{array}{c}
    A\sin\left(\varphi\right)\cos\left(k_i \lambda \right)\sin\left(k_i \varphi \right) \cdot \mathbf{e_{\varphi}} \\
    A\cos\left(\varphi \right)\sin\left(k_i \lambda \right)\sin\left(k_i \varphi \right) \cdot \mathbf{e_{\lambda}}  \\
    0 \cdot \mathbf{e_{z}}
  \end{array}
\right) 
\end{equation}\end{linenomath*}
This approach avoids both forcing intermittency (i.e. semi-annual and annual cycles, as well as eddy bursts) and spatial anisotropy. In addition, the energy injection rate is known analytically from Eq.~\eqref{scalarproduct}, 
\begin{linenomath*}\begin{equation}
\epsilon_i =  \sum_{n=0}^N \sum_{m=-n}^n (\mathbf{u} ,\mathbf{f})_{m,n}.
\end{equation}\end{linenomath*}
The time-averaged injection rate, defined as $\hat{\epsilon_i} = \langle \epsilon_i \rangle_T$, is set to emulate the energy transfer rate $\epsilon$ of our Saturn reference simulation. The forcing magnitude $A=0.22$ m\,s$^{-1}$ leads to a mean energy injection rate that ranges between $5.5 \times 10^{-6} \leq \hat{\epsilon_i} \leq 5.6 \times 10^{-6}$ for all simulations. The injection wavenumber $k_f=56$ project on index $n_i=112$ when the basis functions of spherical harmonics are used. It is chosen to be both small-scale and outside the range where model hypeviscosity affects the flow statistics (i.e. $n_i<n_{\delta}$). \rev{One relatively unconstrained aspect of the dynamics is the strength, and latitudinal extent, of the artificial bottom drag due to the lack of detailed knowledge about exactly where and how the MHD drag acts on the atmospheric circulation. Here, we choose the drag formulation to be the same as for our Saturn reference simulation  \citep{spiga2020} to ensure that differences among the simulations are not caused by differences in this poorly constrained parameter. Investigating the effects of bottom friction that govern flow statistics at large scales is beyond the scope of the present paper.} Then rotation rate is the only varying parameter, which we set to $4\Omega$, $\Omega$, $\Omega/2$ and $\Omega/4$, and each simulation reaches equilibrium over a timescale of one Saturn year 
 (a steady state plateau in energy is reached by 0.2 year, not shown).\\


\subsection{Horizontal velocity and vorticity fields}

\begin{figure}
  \centering
    \includegraphics[width=\textwidth,clip,viewport=15 5 1110 770]{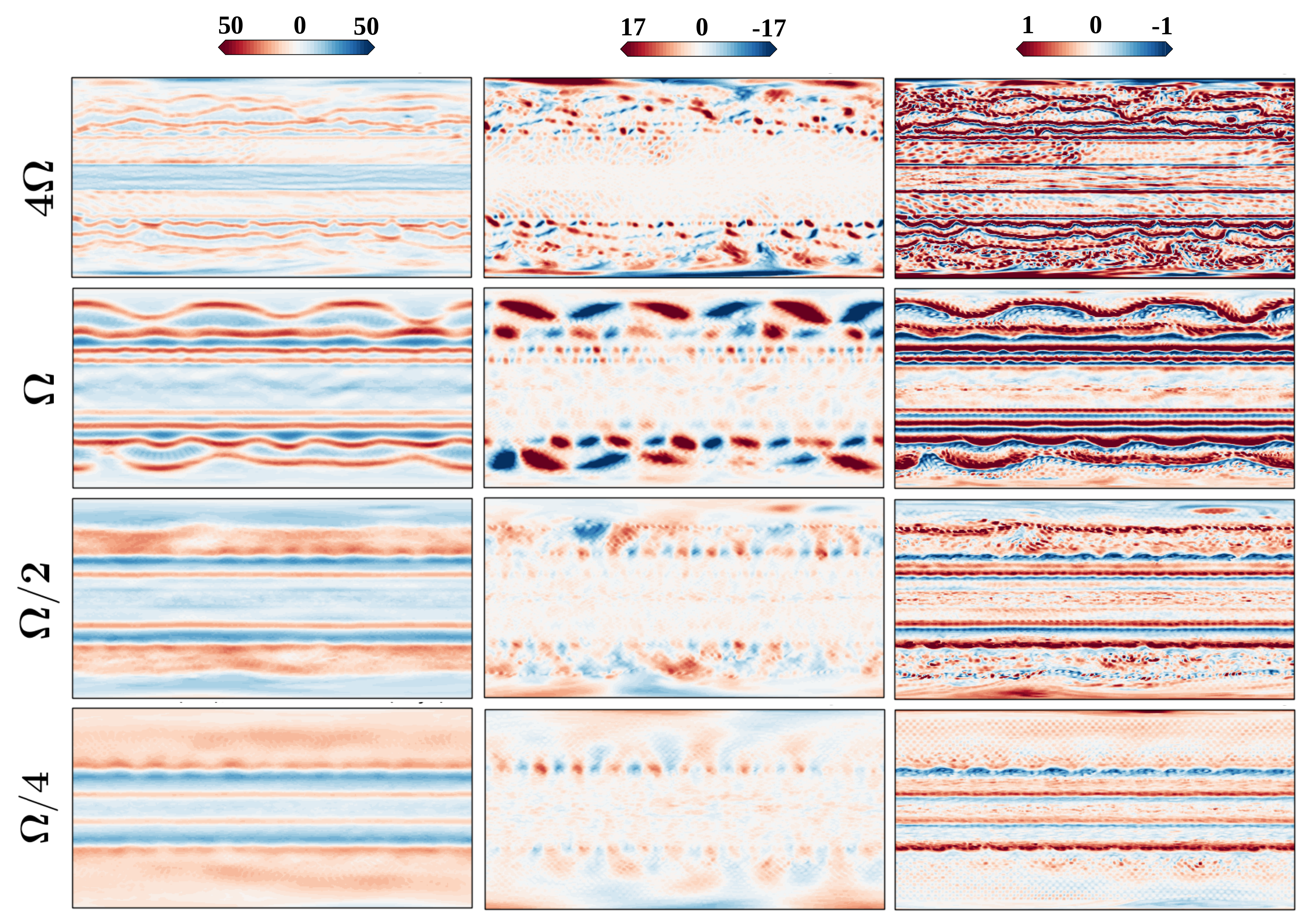}
      \caption{Instantaneous (latitude-longitude) \rev{velocity and vorticity} maps: flow fields are shown for four idealised simulations with varying planetary rotation rate. (Left) zonal velocity (m\,s$^{-1}$), (middle) meridional velocity (m\,s$^{-1}$), and (right) vorticity (s$^{-1}$) at tropospheric depth 94.28 km above 3 bar. \label{Idealized-MAps}}
\end{figure}

Fig.~\ref{Idealized-MAps} shows zonal, meridional and vorticity fields at the last remapped time step and at the same vertical level as the horizontal maps of the Saturn reference simulation in Fig.~\ref{refsimu-HorzSection} (i.e. 94.28\,km above 3\,bar). The simulations are ordered by decreasing rotation rate or, equivalently, by increasing Rossby non-dimensional number. In all four idealized simulations, the combination of low Rossby number with a strong $\beta$-effect leads to velocity and vorticity patterns dominated by multiple eddy-driven zonal jets. The latitudinal jet size decreases with increasing rotation rate, in keeping with Rhines' theory. Increasing rotation is also accompanied by the formation of intense non-axisymmetric modes, which compete with the $m=0$ dominant mode. This is particularly evident at $\Omega$ and $4\Omega$, in which the meridional velocity and vorticity fields are considerably strengthened.

\begin{landscape}
\thispagestyle{empty}
\begin{figure}[p]
\centering
    \includegraphics[width=\columnwidth]{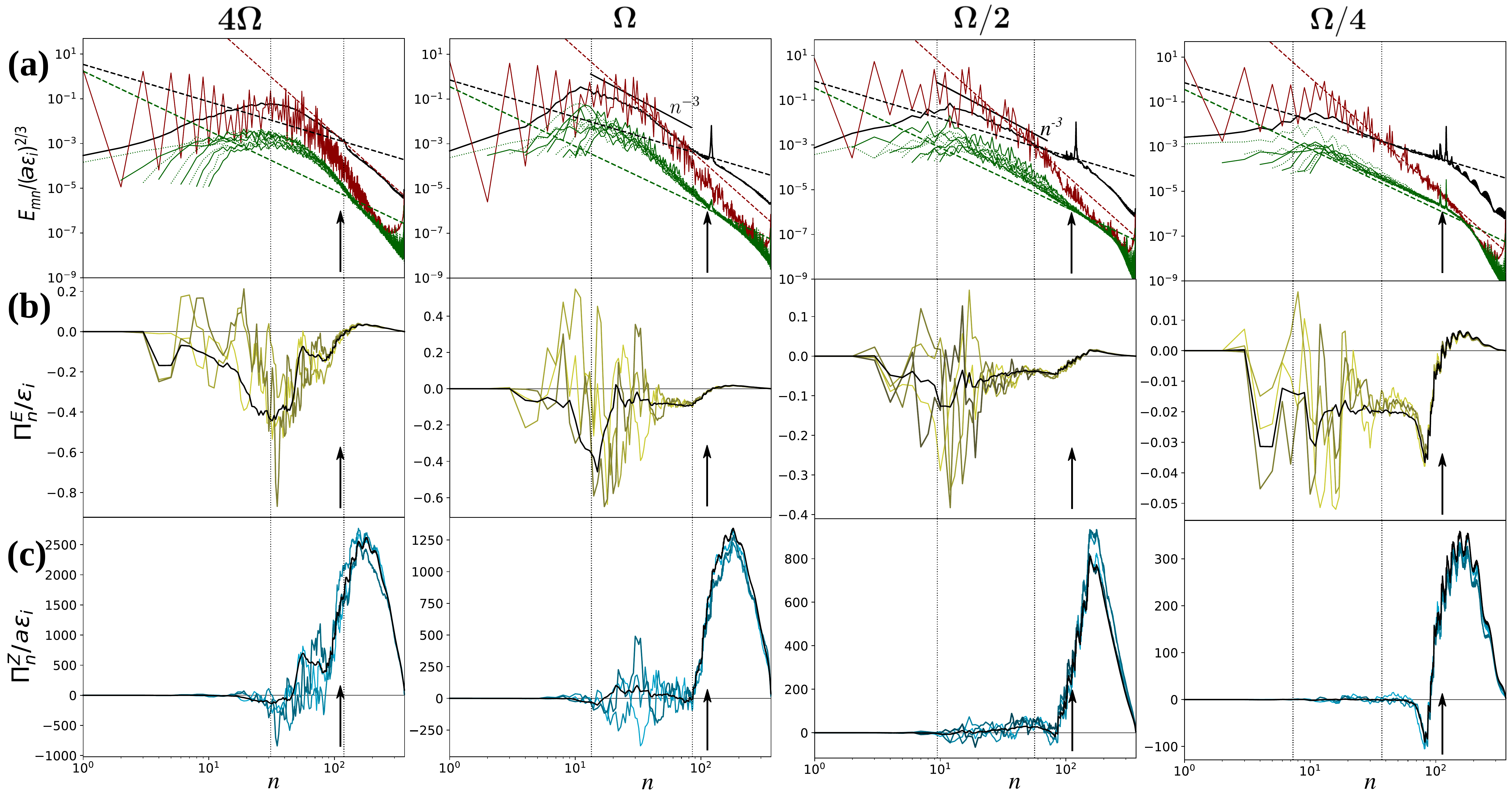}
      \caption{\rev{Spatial kinetic energy spectra as well as energy and enstrophy fluxes as a function of latitudinal index:} spectral quantities are reported for four idealized simulations at various rotation rates. The energy injection index $n_i$ is shown with black vertical arrows. Vertical dashed lines show the Rhines ($n_R$) and transitional ($n_{\beta}$) indices from left to right in each panel. Vertical arrows correspond to the injection index $n_i=112$. (a) Time-averaged normalized kinetic energy distribution showing zonal energy $E_Z(n)/(\epsilon_i a)^{2/3}$ (red), residuals $E_R(n)/(\epsilon_i a)^{2/3}$ (black), and modal spectra up to $m=10$, $\mathscr{E}(n,m)/(\epsilon_i a)^{2/3}$ (green), as well as their theoretical predictions (Eqs.~\ref{KESa}, \ref{KESb}, and \ref{eq:modalspectrum}). (b) Time-averaged (black) and instantaneous (yellow) energy fluxes $\Pi^E/\epsilon_i$. (c) Time-averaged (black) and instantaneous (blue) enstrophy fluxes $\Pi^Z/(\epsilon_i^{-1} a^2)$.\label{Idealized-spectra-complet}}
\end{figure}
\end{landscape}

\subsection{Energy spectra of the wind fields}

By computing spherical harmonic decompositions of the wind fields, we show in Fig.~\ref{Idealized-spectra-complet} the time-averaged residual, zonal and modal spectra for each rotation rate. 
Zonostrophic turbulence is a useful framework for interpreting the statistical energy distributions from these simulations. In all simulations (except $4\Omega$ that slightly departs from predictions) the spectra display a self-similar form with a well delimited zonostrophic inertial range in which zonal energy dominates residuals with a magnitude proportional to $\beta^2$ (Eq.~\ref{KESa}). The Rhines index locates quite effectively the most energetic scale, above which energy systematically decreases. Table~\ref{tab:idealised} lists the indices and non-dimensional numbers estimated from our spectral analysis for each simulation.

The simulation at rotation $\Omega/4$ has similar statistical properties to numerical studies that model 2D-turbulent flow on the surface of a rotating sphere \citep{sukoriansky02,sukoriansky07,galperin10}. We retrieve the two distinct spectral domains shown in Fig.~\ref{theory}: the zonostrophic inertial range for $n_{R} < n < n_{\beta}$, and the residual-dominated range for $n_{\beta} < n < n_i$.  The residual spectrum follows the KBK $n^{-5/3}$ slope, whose magnitude at the injection scale is estimated to be $\epsilon \simeq 3 \times 10^{-7}$ m$^2$\,s$^{-3}$. 
No energy is injected at the jet scale, ensuring a smooth zonal spectrum and a peaked residual spectrum at injection. The ``zigzag'' features of the zonal spectrum at small indices $n$ are due to the hemispheric symmetry of the zonal structure, which projects onto the even indices. The modal spectrum shows equally distributed modes apart from the axisymmetric mode $m=0$, which dominates the dynamics with a magnitude equal to $C_Z\beta^2$ (see Eq.~\ref{KESa}). Finally, all modes intersect at $n_z$ and the zonostrophy index, the Rossby index and the associated Rossby number are respectively $R_{\beta} = 5.1$, $n_{Ro}=8\times 10^4$, and $Ro=9\times 10^{-3}$.

\begin{table}[tb]
\centering
 \begin{tabular}{|ccccccccc|} 
  \hline
  &  $n_Z$ & $n_{\beta}$ &  $n_R$ & $R_{\beta}$& $\epsilon$ (m$^2$\,s$^{-3}$)  & $n_{Ro}$ & Ro & $E_{rot}/E_{div}$\\ [0.5ex] 
 \hline
 SRS  & 449 &  58 & 11 & 5.3 &$2 \times 10^{-6}$ &$2.7 \times 10^{5}$ & $4 \times 10^{-3}$ & $200$\\
   \hline
 $4\Omega$  & 1238   &  118 & 31  & 3.9 & $6 \times 10^{-6}$ & $2 \times 10^{6}$  & $5 \times 10^{-4}$ & $100$ \\
 $\Omega$  & 773 &  85 & 13 & 6.3 &$3 \times 10^{-7}$&$7 \times 10^{5}$ & $3 \times 10^{-3}$ & $1000$\\
  $\Omega/2$  & 427 & 56 & 9.5 & 5.9 &$3 \times 10^{-7}$& $2 \times 10^{5}$ & $5 \times 10^{-3}$  & $1000$\\
  $\Omega/4$  & 235 & 37 & 7.3 & 5.1 &$3 \times 10^{-7}$& $8 \times 10^{4}$ & $9 \times 10^{-3}$ & $200$\\
 \hline
\end{tabular}
  \caption{Summary of energy diagnostic quantities, for comparison between our Saturn reference simulation (SRS) and the idealized simulations. } \label{tab:idealised}
\end{table}

Considering simulation $\Omega/4$ as a benchmark that reproduces the spectral features of zonostrophic turbulence, we now explore higher rotation rates to frame our Saturn reference simulation within an idealized 3D-turbulent set-up. At $\Omega/2$, the zonostrophy index and Rossby index increase to $R_{\beta} = 5.9$ and $n_{Ro}=2\times 10^5$ ($Ro=9\times 10^{-3}$) respectively. These non-dimensional quantities are closest among the idealized simulations to those obtained for the Saturn reference simulation (see Table~\ref{tab:idealised}). Indeed, the lower rotation rate $\Omega/2$ is balanced by the lowest kinetic energy injection rate $\epsilon$, leading to a similar Rossby number and strength of the zonostrophic regime. Similarly to SRS, the residual spectrum features a $n^{-5/3}$ slope that exists between the injection scale and the transitional scale (i.e. in the residual-dominated range $n_{\beta} < n <n_i$) and a $n^{-3}$ slope within the zonostrophic inertial range \rev{that can be related to the growth of non-axisymmetric modes in the same fashion as for SRS Fig.~\ref{Spectra-Transitory}.} \rev{Also, the steepening to a $n^{-3}$ slope in the residual spectrum might stem from the large value of the friction index \citep[$n_{fr} \sim n_R$ in the zonostrophic regime, see][]{sukoriansky07} leading to erosion and eventual destruction of KBK statistical laws at large scale, as shown in \cite{sukoriansky99}. If such a $n^{-3}$ spectral slope is not retrieved in direct observations of Saturn's flow ($n^{-3}$ slope being a common feature of the SRS and its analogue ideal simulation at $\Omega/2$), then it indicates that bottom friction might have to be implemented in a more elaborate way in our GCM.}\\

Moving to rotation rate $\Omega$, the zonostrophic and Rossby indices rise to  $R_{\beta} = 6.3$ and $n_{Ro}=7\times10^5$ ($Ro=3\times 10^{-3}$) respectively, and the dynamical confinement by the jets is reinforced. Again, the residual energy spectrum is substantially steepened and the modal spectrum now features energetic zonal modes that pile up within the zonostrophic inertial range (predominantly with $m=1$ and $m=4$). These modes, easily discernible in the velocity maps (Fig.~\ref{Idealized-MAps}), break the flow's zonal axisymmetry, leading to a dynamical regime dominated by strong zonal waves. 
Neither the classical KBK energy cascade nor the zonostrophic regime of \cite{sukoriansky02} are fully applicable to this flow regime. By increasing the rotation rate again to $4\Omega$ we suppress the residual-dominated inertial range ($n_{\beta}<n<n_i$), 
the transitional index $n_{\beta}$ nearly coincides with the injection scale and the resulting spectra are curved. 
The lack of scale separation forces energy accumulation within a restricted spectral range bordered by the Rhines and injection scales. \rev{The associated low zonostrophy index $R_{\beta} = 3.9$ suggests that the dynamics at $4\Omega$ are dominated by frictional effects that arrest the inverse cascade at relatively small scale compare to other simulations.} Consequently, all modes have a spectrum steeper than $n^{-5}$, and the regime of zonostrophic turbulence is therefore not applicable to this simulation.


The tendency of the residual spectrum to develop a maximum at small total wavenumber, with a steeper slope at the adjacent smaller scales ($\approx n^{-3}$), is shared by all simulations with rotation rates larger than $\Omega/4$, including the SRS. \rev{It has been suggested that the bumpy residual spectrum ($\approx n^{-3}$ slope) can be related to the breaking of large scale zonal jets that are barotropically unstable and produce non-axisymmetric zonal modes.} 
Thus, the presence of \rev{these} intense zonal modes, inducing a $\approx n^{-3}$ slope, is a consequence of the dynamical regime rather than a product of the intermittent-anisotropic forcing specific to the Saturn reference simulation. 
One possible argument is that non-linear wave interactions play an important role in 3D turbulent models \citep[see, e.g.,][]{yarom14} and induce the formation of strong zonal modes that steepen the spectral signature. This was indeed not observed in previous 2D models where depth invariance was assumed and not produced spontaneously by the dynamics. \rev{Another possible explanation is that bottom friction distorts the statistics of the inverse cascade at large scale \citep{sukoriansky99}, but testing this hypothesis is beyond the scope of this work.} In the large-scale range $n<n_R$, the dominant long-lived modes $m=2$ and $m=3$ that are observed in the Saturn reference simulation are absent in all idealized cases. As a consequence, they are presumably related to the baroclinic conditions specific to the SRS. 

Finally, we note that rotation rate only weakly impacts the $n^{-5}$ spectral signature of the zonal spectrum for all simulations (except at $4 \Omega$, where the zonostrophic regime is not applicable). This is also true for residual spectra at scales smaller than the injection scale, where the spectral slope is invariably close to $n^{-3}$. This is consistent with the associated downscale enstrophy cascade at indices $n>n_i$, but it could instead, and more logically, result from the hypeviscosity damping that prevails at small scales. 

\subsection{Spectral fluxes}

Figure~\ref{Idealized-spectra-complet}(b,c) shows time-averaged spectral fluxes as well as instantaneous spectral fluxes randomly chosen from within the steady state period. In our benchmark simulation $\Omega/4$, we retrieve the double cascade scenario of  2D-turbulent flows, characterized by a downscale cascade of enstrophy at small scales ($n \gtrsim n_i$) and an upscale energy cascade forming an extended negative plateau at large scale ($n \lesssim n_i$). Looking at the instantaneous spectral quantities (colour gradient), we find strong deviations from the mean fluxes.  If the enstrophy flux is only weakly varying around its time average, the energy flux displays strong variations, specifically intensified beyond the transitional index, i.e. for $n \lesssim  n_{\beta}$. Again, the zonostrophic and residual-dominated subranges feature different dynamics. The residual-dominated range (i.e. $n_{\beta} \lesssim n \lesssim n_i $) is characterized by a stably negative energy flux that is the counterpart of the entrenched $n^{-5/3}$ slope in the residual spectrum. 

On the other hand, the zonostrophic inertial range is marked by a weak energy bump in the residual spectrum, and the associated energy fluxes strongly oscillate around the mean state, sometimes becoming positive. This implies intermittent periods of upscale and downscale energy transfer, which occurs only beyond the transitional index $n_{\beta}$. Strongly varying in space and time, these direct and inverse energy cascades are not visible when looking only at averaged spectra. Moreover, as they are present in both our idealized simulations and in the SRS, the coexistence of positive and negative energy transfers at large scale does not appear to be related to the nature of the energy forcing.  \\

Higher rotation rates show the same global tendency except that the lack of scale separation considerably reduces the residual-dominated range ($n_{\beta} \lesssim n \lesssim n_i$) within which the $n^{-5/3}$ slope develops. Due to the increasing efficiency of dynamical confinement in latitude at larger Rossby index (and hence lower Rossby number), the upscale energy flux increases by an order of magnitude between $\Omega/4$ and $\Omega$. Note that energy fluxes are all normalized by the energy injection rate $\Pi^E /\hat{\epsilon_i}$ so the flux shows the proportion of energy transferred with respect to the energy that was injected. Deviations from the mean fluxes also increase with increasing rotation rate, and show a succession of strengthened upscale and downscale energy transfers. The mean energy flux remains upscale, however, as it is necessary to constantly sustain the large-scale jets. Enstrophy is also strongly impacted by the intense activity of non-axisymmetric modes and, at each downscale energy peak, corresponds to a direct enstrophy cascade. \rev{It is also interesting that the $n^{-3}$ slope in the energetic spectra has positive enstrophy flux (on average) within the zonostrophic inertial range. This spectral signature ($n^{-3}$) of a downscale enstrophy cascade can be related to energy accumulation and enstrophy production at large scales. In summary, for all simulations, the amplitude of the residual spectrum at large scales ($n_R < n < n_{\beta}$) increases with rotation rate and also with the magnitudes of both the direct enstrophy cascade and the inverse energy cascade, featuring a bumpy spectrum with a nearly $n^{-3}$ slope. }


\section{Conclusion \label{conclu}}

We have presented a phenomenological study of dynamical energy and vorticity transfers between different horizontal scales in a 3D-turbulent Global Climate Model. 
 This constitutes a coherent and advanced statistical analysis of the planetary flow implemented by \citet{spiga2020} using the new dynamical core DYNAMICO, with the unprecedented combination of fine horizontal and vertical resolution, and the incorporation of atmospheric physical processes such as radiative transfer \citep{guerlet14}, emulating Saturn's troposphere and stratosphere. 
 
 We found spectral evidence that an inverse energy cascade occurs under rapid rotation and shallow-depth confinement in a 3D-turbulent atmospheric layer. We recover several statistical properties observed in more classical regimes of forced rotating turbulence (specifically the inverse energy cascade due to flow barotropization, \cite{pouquet2013,deusebio14,yarom14}; \rev{and the dual cascade in forced rotating turbulence was analyzed by \cite{sukoriansky16} analytically}) and we show some statistics specific to Saturn's atmospheric dynamics. The growth of intense baroclinic eddies at small planetary scales and the coupled action of low Rossby and a strong $\beta$-effect induces a zonal spectral anisotropy that implies the presence of a zonostrophic turbulent regime \citep{sukoriansky02}. A set of zonal jets emerges as the dominant flow structure, and the flow's statistical properties follow the double cascade scenario reminiscent of the 2D turbulence paradigm \citep{Kraichnan67}. Scaling analysis reveals a turbulent regime in our Saturn reference simulation that is similar to that found in Jupiter's atmosphere \citep{galperin14, young17}. Energy accumulation into the barotropic mode is also a feature shared by our Saturn simulation and Jupiter as shown in recent observations from Juno mission \citep{kaspi18} and in a simulation of a Jupiter-like atmosphere by \cite{liu15}. Non-axisymmetric modes with $m=2$ and $m=3$, which are specific to Saturn's atmospheric dynamics, are also retrieved in our spectral analysis \citep{spiga2020}. \rev{However, although GCM simulations have advanced a lot since the pioneering shallow models of \cite{cho96}, it is clear that shallow models still lack a naturally arising prograde equatorial jet, whereas they arise generically in deep convection models \citep{heimpel05}. GCMs are missing a fundamental geometrical constraint that leads to the dominant scaling in the equatorial region. 
 However, at higher latitudes, only models that are restricted to a thin shell (inner-to-outer radius ratio $> 0.9$) appear capable of reproducing the multiple, mid-latitude jets, which are too few and too broad in simulations using deep convective models \citep{vasavada05}. }\\


 
By running idealized simulations in a simplified GCM, with a well-controlled energy forcing and for various rotation rates, we reproduce flows with a varying degree of agreement with the zonostrophic regime, framing the dynamics of our Saturn reference simulation. A set of four idealized simulations allowed us to clearly identify three inertial ranges with distinctive dynamics and specific spectral signatures:
\begin{itemize}
    \item A residual-dominated range ($n_{\beta} < n < n_i$) characterized by a KBK $n^{-5/3}$ slope in the residual kinetic energy spectrum. The energy flux is stably negative and the enstrophy flux is nearly zero.
    \item A zonostrophic inertial range  ($n_{R} < n < n_{\beta}$), marked by a steeper $\approx n^{-3}$ slope with a clear maximum in the residual energy spectrum, as well as a succession of erratic upscale and downscale energy and enstrophy transfers that dominate the mean dynamics. It has been suggested that the steepening of the residual energy spectrum results from the pile-up of non-axisymmetric modes at the jet scale \rev{that are produced by jet instability (likely barotropic instability)}. Such an energy \rev{accumulation} at large scales is in turn likely responsible for the succession of positive and negative spectral fluxes at large scales. \rev{It has also been mentioned that bottom friction possibly affects the dynamics at large scale. It might, for example, be partially responsible for the energetic bump ($n^{-3}$) in the residual spectra. If such a feature is not retrieved from direct observations of Saturn's flow, then GCMs might improve their predictions by implementing a more elaborate bottom drag formulation \citep{sukoriansky99}.}
    \item A large scale range, for indices $n < n_R$, dominated by friction and marked by an energy decrease. Energy transfers are nonetheless efficient up to this large scale range.
\end{itemize}

These three spectral ranges are observed in our Saturn reference simulation, assuming that kinetic energy is injected near the Rossby deformation radius. Only the dynamics within the large scale range differ from the idealized simulations, where it is marked by the presence of the two long-lived modes $m=2$ and $m=3$. The main differences between the Saturn reference simulation and our idealized simulations are its intermittent-anisotropic forcing resulting from the growth of stratospheric baroclinic instabilities over seasonal cycles. As a consequence, while the non-dimensional quantities computed from the SRS                                                                  are similar to those from our $\Omega/2$ idealized simulation, the ratio between rotational and divergent energy is significantly different, with $E_{rot}/E_{div} \approx 100$ for the SRS                                                                  compared with $\approx 1000$ for the idealized case (see Table~\ref{tab:idealised}). The periodic energy injection in the topmost atmospheric layers requires a continuous adjustment of the dynamics that manifest as a strengthened role of the divergent velocity component, leading to the specific statistical features of our Saturn reference simulation. \\

In this study, we used only the rotational (non-divergent) part of the horizontal velocity field when calculating the spectral fluxes.
This spectral flux based on the 2D vorticity equation is primarily responsible for the upscale cascade while the divergent part contribution increases in the downscale cascade range \citep{augier13}. In our configuration the flow is highly barotropized, the upscale cascade covers a wide spectral domain, and the energy in the rotational part of the flow is much larger than in the divergent part, by a factor $\approx 100$ to $1000$ depending on the simulation. In addition, the downscale cascade is presumably poorly resolved due to hypeviscosity effect at small scales. Nevertheless, the divergent part of the spectral flux should be examined more closely in the future and might reveal new statistical features in our Saturn reference simulation. Furthermore, the implementation of atmospheric moist convection as an additional kinetic energy source will also significantly impact the dynamics, with its own spectral signature.

A last (but not least) limitation of our approach is the horizontal and vertical resolution. Baroclinic eddies can develop but are distorted by hypeviscosity. Increasing the horizontal resolution from eddy-permitting to eddy-resolving would likely improve the representation of the global energy budget. Ultimately, the strength of the upscale energy transfer is decisive in setting the large scale non-axisymmetric dynamics, marked by the presence of zonal modes. Dynamical confinement in latitude being fixed by Saturn's rotation rate, the strength of the upscale cascade might vary with vertical resolution and/or depth extension through geometric confinement. However, the exploration of various layer depths is beyond the scope of the present study. 

\section*{Acknowledgments}

The authors 
acknowledge exceptional computing support from 
Grand \'{E}quipement National de Calcul Intensif (GENCI)
and
Centre Informatique National de l'Enseignement Sup\'{e}rieur (CINES).
All the simulations presented in this paper were carried out on the \emph{Occigen} cluster hosted at CINES.
This work was granted access to the High-Performance Computing (HPC) resources 
of CINES under the
allocations A001-0107548, A003-0107548, A004-0110391 made by GENCI.
The authors
acknowledge funding from 
Agence Nationale de la Recherche (ANR),
project HEAT ANR-14-CE23-0010
and project EMERGIANT ANR-17-CE31-0007.
This project has received funding from the European Union's Horizon 2020 research and innovation programme under the Marie Sklodowska-Curie grant agreement No 797012.
Fruitful discussions with 
Sandrine Guerlet, 
Ehouarn Millour,
Thomas Dubos, 
Fr{\'e}d{\'e}ric Hourdin
and Alexandre Boissinot
from our team helped refine some discussions in the paper.






\bibliographystyle{elsarticle-harv-ed}
\bibliography{biblio}



\end{document}